\documentclass[aps,prb,twocolumn,superscriptaddress]{revtex4-2}

\usepackage{amsmath}
\usepackage{amssymb}
\usepackage{graphicx}
\usepackage{xcolor}
\usepackage{bm}
\usepackage{hyperref}

\begin{document}

\title{Summing Real Time Feynman Paths of Lattice Polaron with Matrix Product States}

\date{\today}

\author{Qi Gao}
\affiliation{Institute of Physics, Chinese Academy of Sciences, Beijing 100190, China}
\affiliation{University of Chinese Academy of Sciences, Beijing 100049, China}

\author{Yuan Wan}
\email{yuan.wan@iphy.ac.cn}
\affiliation{Institute of Physics, Chinese Academy of Sciences, Beijing 100190, China}
\affiliation{Songshan Lake Materials Laboratory, Dongguan, Guangdong 523808, China}

\begin{abstract}
We study numerically the real time dynamics of lattice polarons by combining the Feynman path integral and the matrix product state (MPS) approach. By constructing and solving a flow equation, we show that the integrand, viewed as a multivariable function of polaron world line parameters, can be compressed as a low bond dimension MPS, thereby allowing for efficient evaluation of various dynamical observables. We establish the effectiveness of our method by benchmarking the calculated polaron spectral function in one dimension against available results. We further demonstrate its potential by presenting the polaron spectral function in two dimensions and simulating polaron diffusion in both one and two dimensions.
\end{abstract}

\maketitle

\section{Introduction \label{sec:intro}}

Efficient simulation of quantum many-body dynamics is a long-standing problem in condensed matter physics~\cite{Park1986,Aoki2014,Paeckel2019,Schmitt2020}. The difficulties with this task are amply illustrated by a prototypical $N+1$ body system, a polaron, which consists of an electron interacting with phonons. Instrumental in the historical development of many-body physics~\cite{Landau1933,Landau1948,Feynman1955,Holstein1959}, it has received renewed attention~\cite{Ciuchi1997,Fratini2003,Berciu2006,Goodvin2006,Pandey2022,Mitric2022} owing to its rich dynamical properties~\cite{Vidmar2011,Golez2012} and potential connection to high-temperature superconductors~\cite{Alexandrov1981,Alexandrov1986,Chakraverty1998,Sous2018,Fetherolf2020,Zhang2023} as well as solar cells~\cite{Zhu2015,Zhu2016,Fu2019,Ghosh2020}.

Despite decades of progress~\cite{Alexandrov2010}, simulating the real time dynamics of polaron still faces obstacles. On one hand, Monte Carlo simulation in real time often encounters dynamical signs~\cite{Loh1990,Cohen2015}. While the sign-free imaginary time Monte Carlo is efficient and accurate in determining low energy properties of polaron~\cite{Prokofev1998,Mishchenko2000}, accessing its spectral functions requires analytic continuation~\cite{Mishchenko2003,DeFilippis2006,Goodvin2011,Mishchenko2015}. On the other hand, variational approaches using various wave function ans\"{a}tze~\cite{Romero1998,Wellein1998,Zhang1998,Zhang1999,Bonca1999,Barisic2004,Dolgirev2021,Bonca2021} including the matrix product state (MPS)~\cite{Schollwock2011,Jeckelmann1998,Zhao2023} are successful in one dimension. Yet, going to higher spatial dimensions remains challenging~\cite{Hohenadler2003,Ashida2018,Shi2018,Zhao2024}. In particular, the infinite local Hilbert space dimension of phonons needs proper truncation~\cite{Jeckelmann1998,Zhang1998}.

\begin{figure}
\includegraphics[width = \columnwidth]{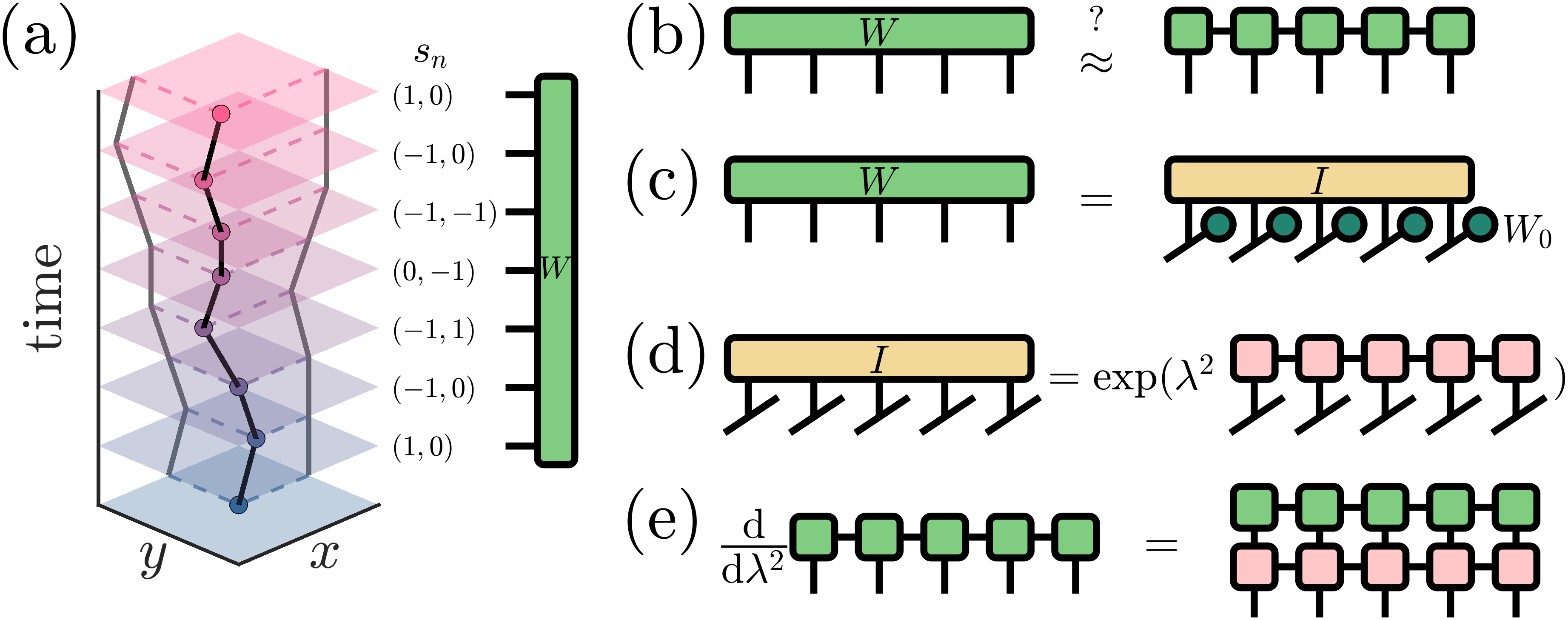}
\caption{MPS summation of real time Feynman paths. (a) World line of a polaron is parametrized by its displacements $\{s\}$ between successive times. The amplitude $W$ as a function of $\{s\}$ may be viewed as a high dimensional tensor. (b) $W$ can be efficiently summed provided that it is well approximated by a low bond dimension MPS. (c) $W$ is given by the free amplitude $W_0$ dressed by phonon influence functional $I$, which, in turn,  is the exponential of a low bond dimension matrix product operator (d). (e) $W$ satisfies an ordinary differential equation with respect to electron-phonon coupling $\lambda^2$. Integrating it from $\lambda^2=0$ yields the MPS approximation to $W$.}
\label{fig:sketch}
\end{figure}

In this work, we tackle the polaron dynamics problem by taking advantage of both path integral and MPS approaches. Focusing on the lattice polaron, we show that its real time Feynman path integral can be efficiently evaluated by casting it in MPS form (Fig.~\ref{fig:sketch}a\&b). Recall a path integral is essentially a high dimensional sum:
\begin{align}
Q = \sum_{\{s\}} W(s_M,\cdots s_2,s_1),
\end{align}
where $Q$ is some physical quantity of interest. The labels $\{s\}$ parametrize the world line, and $W$ is its amplitude. The main idea behind our method is to approximate $W$ as a low bond dimension MPS, 
\begin{align}
\label{eq:mps_apprx}
W(s_M,\cdots s_2,s_1) \approx W_M (s_M)\cdots W_2(s_2) W_1(s_1),
\end{align}
where $W_1(s_1)$, $W_2(s_2)$, $\cdots W_M(s_M)$ are $s$-dependent matrices of size $1\times \chi_{M-1}$, $\cdots$, $\chi_2\times\chi_1$, $ \chi_1 \times 1$. Here, $\chi_1$, $\chi_2$, $\cdots \chi_{M-1}$ are MPS bond dimensions. By virtue of this approximation, the $M$-dimensional summation is reduced to $M$ independent, one-dimensional summations:
\begin{align}
Q &\approx \sum_{s_M} W_M (s_M)\cdots  \sum_{s_2} W_2(s_2)\sum_{s_1}W_1(s_1)
\nonumber\\
& = W_M \cdots W_2 W_1.
\end{align}
$W_j \equiv \sum_{s_j}W_j(s_j)$. In the first line, we sum each and every $s_j$, producing the matrices $W_1$, $W_2$, $\cdots W_M$. The computational cost of this step is $Md\chi^2$, where $d$ is the range of the world line parameter $s_i$, or the physical dimension in MPS terminology, and $\chi$ is the maximal bond dimension. In the second line, we evaluate the matrix product with the computational cost on the order of $M\chi^2$. Note the cost for summing $W(\{s\})$ by brute force  is exponentially large ($d^M$). Crucially, this scheme for high dimensional summation can  tolerate rapid complex phase fluctuations in amplitude $W(\{s\})$ that plague conventional Monte Carlo summation.

This method shares similar sprit with the tensor train method for high dimensional integration~\cite{Oseledets2010,Oseledets2011,Dolgov2020,NunezFernandez2022,NunezFernandez2025}, as well as the tensor network influence functional approach to spatially local open quantum systems~\cite{Strathearn2018,Joergensen2019,Cygorek2022,Chen2024} and one-dimensional systems~\cite{Banuls2009,Huang2014,Lerose2021,Fux2023}, but also features perspective and formulation that are unique to polarons. The most difficult step of our summation scheme is finding the MPS approximation Eq.~\eqref{eq:mps_apprx} of the multivariable function $W(\{s\})$. For polaron problems, the analytic expression of this function is well known~\cite{Feynman1955}. However, it is not given in a form that can be readily compressed as low bond dimension MPS. We overcome this problem by deriving a differential equation for $W(\{s\})$, dubbed the flow equation, which admits a variational solution in the form of low bond dimension MPS (Fig.~\ref{fig:sketch}e). Said differently, solving the flow equation with the MPS ans\"{a}tz produces the desired MPS approximation Eq.~\eqref{eq:mps_apprx}.

Note the problem at hand is quite different from the more familiar variational approach that utilizes MPS~\cite{Schollwock2011,Jeckelmann1998,Zhao2023}. There, the ground or excited state wave functions are \emph{unknown}; the MPS serves as a variational ans\"{a}tz for these wave functions. By contrast, the function that we wish to compress, $W(\{s\})$, already has a closed, analytic expression. We seek its MPS approximation Eq.~\eqref{eq:mps_apprx} to facilitate the summation.

As we shall see, our method is sign tolerant, applicable to higher spatial dimensions, free from truncation in phonon Hilbert space, and admits various generalizations. The rest of this work is organized as follows. We present our method in Sec.~\ref{sec:method}. In Sec.~\ref{sec:results}, we establish its effectiveness by benchmarking the polaron spectral function, calculated through real time evolution, against available results in one dimension. We further demonstrate its potential by computing the spectral function in two dimensions and taking a first pass at simulating a nonequilibrium diffusion process. Finally, we provide a tentative explanation for the effectiveness of our method, and the potential extensions of our method in Sec.~\ref{sec:discussion}.

\section{Method \label{sec:method}}

In this section, we give a detailed description of our method. We set the stage with a one-dimensional polaron model in Sec.~\ref{sec:setup}. In Sec.~\ref{sec:path_integral}, we derive a path integral expression for the electron propagator. We define the phonon influence functional along the way. In Sec.~\ref{sec:exponentiation}, we derive a key result of this work, namely the phonon influence functional is the exponential of an MPS with low bond dimension. Building on this result, in Sec.~\ref{sec:compression}, we develop a numerical method that can compress the path integral amplitude as low bond dimension MPS.

\subsection{Problem setup \label{sec:setup}}

We use a one-dimensional polaron model to illustrate our method~\cite{Marchand2013}. The Hamiltonian is given by:
\begin{subequations}
\label{eq:hamiltonian}
\begin{align}
H = K+V.
\end{align}
$K$ describes electron hopping in a chain of $N$ sites subject to periodic boundary condition:
\begin{align}
K = -\sum_{i} (c^\dagger_i c^{\phantom\dagger}_{i+1} + c^\dagger_{i+1} c^{\phantom\dagger}_i).
\end{align}
$c_i$ ($c^\dagger_i$) annihilates (creates) a spinless electron on site $i$. The hopping amplitude is set to $1$ by rescaling the unit of energy. $V$ describes the phonons and their coupling to electrons:
\begin{align}
V = \sum_{q} [ \omega_{q} b^\dagger_{q} b^{\phantom\dagger}_{q}-\lambda f_{q}n_{q} (b_{q} + b^\dagger_{-q})].
\end{align}
\end{subequations}
$b_{q}$ ($b^\dagger_{q}$) annihilates (creates) a phonon with wave vector $q$. $\omega_{q}$ is the dispersion relation. The electron-phonon coupling constant $\lambda>0$. $f_{q}$ is the form factor, which satisfies $f_q = f^\ast_{-q}$. $n_{q}$ is the Fourier transform of electron density operator, given by $n_q = \sum_{k} c^\dagger_{k+q} c^{\phantom\dagger}_{k}/\sqrt{N}$. Note $H$ conserves the total number of electrons $N_e = \sum_i c^\dagger_i c^{\phantom\dagger}_i$.

The quantity of interest is the zero temperature electron propagator in the momentum space:
\begin{align}
iG^R(k,t) = \langle 0| c^{\phantom\dagger}_k e^{-iHt} c^\dagger_k |0\rangle.
\label{eq:G_def}
\end{align}
$|0\rangle$ refers to the ground state of Eq.~\eqref{eq:hamiltonian} in the zero-electron sector ($N_e=0$). As $c^\dagger_k$ creates an electron with momentum $k$, the ensuing time evolution occurs in the one-electron sector ($N_e=1$). We always assume the time $t>0$ in this work.

\subsection{Path integral \label{sec:path_integral}}

In the first step, we derive a real time path integral representation for the electron propagator Eq.~\eqref{eq:G_def}. To this end, we use the second order Suzuki-Trotter decomposition:
\begin{align}
iG^R(k,t) = \langle 0| c^{\phantom\dagger}_k  (e^{-i\frac{\epsilon}{2}V}e^{-i\epsilon K} e^{-i\frac{\epsilon}{2} V})^M c^\dagger_k |0\rangle.
\label{eq:trotter}
\end{align}
$M$ is the total number of time slices. The Trotter time $\epsilon = t/M$. The Trotter error is $O(\epsilon^3)$.  Since the time evolution occurs in the one-electron sector, we insert in Eq.~\eqref{eq:trotter} partial resolution of identity by electron position eigenstates:
\begin{align}
iG^R(k,t) = \sum_{\{x\}} e^{ik x_M}  I(\{x\})W_0(\{x\}).
\label{eq:path_integral_x}
\end{align}
$\{x\} = \{x_{M},\cdots x_1,x_0\}$ are electron position at successive times. We impose the boundary condition $x_0 = 0$, whereas the summation over $x_{n\ge1}$ is unrestricted. The phase factor $\exp(ikx_M)$ is due to the Fourier transform from real space to momentum space. $W_0$ is the amplitude for a free electron:
\begin{align}
W_0(\{x\}) = \prod^M_{j=1}\langle x_{j}|e^{-i\epsilon K}|x_{j-1}\rangle.
\end{align}
$I$ is the phonon influence functional:
\begin{align}
I(\{x\}) = \langle \phi |e^{-i\frac{\epsilon}{2}V(x_M)}\cdots e^{-i\epsilon V(x_1)}e^{-i\frac{\epsilon}{2}V(x_0)}|\phi\rangle,
\label{eq:I_def_x}
\end{align}
where $|\phi\rangle$ is the phonon vacuum characterized by $b_q|\phi\rangle = 0$. Here, we have used the fact that $V$ commutes with electron coordinate $x$. $V(x)$ is the phonon Hamiltonian when the electron is located at $x$:
\begin{align}
V(x) = \sum_q (\omega_q b^\dagger_q b^{\phantom\dagger}_q - \frac{\lambda f_q e^{iqx}}{\sqrt{N}} b_q - \frac{\lambda f^\ast_{q} e^{-iqx}}{\sqrt{N}} b^\dagger_{q}).
\label{eq:V_x}
\end{align}
Note the product of operators that appear in Eq.~\eqref{eq:I_def_x} act only on the phonon Hilbert space.

In the next step, we perform a change of variable from electron position $x_j$ to electron displacement $s_j = x_j-x_{j-1}$:
\begin{align}
\label{eq:path_integral_s}
iG^R(k,t) = \sum_{\{s\}} (\prod^M_{j=1}e^{iks_j}) W_0(\{s\}) I(\{s\}).
\end{align}
$\{s\} = \{s_M,\cdots s_2,s_1\}$ are the sequence of electron displacements. The first term on the right hand side is the phase factor due to Fourier transform, $\exp(ikx_M)$; we have used the fact $x_M = \sum_j s_j$.  The free amplitude $W_0$ now factorizes:
\begin{subequations}
\label{eq:A0_s}
\begin{align}
W_0(\{s\}) =  \prod^M_{j=1}f(s_j),
\end{align}
where the function $f(s)$ is defined as:
\begin{align}
f(s) \equiv \langle x+s|e^{-i\epsilon K} |x\rangle = 
\left\{\begin{array}{cc}
 1-\epsilon^2 & (s=0) \\
  i\epsilon  & (s = \pm 1) \\
-\epsilon^2/2 & (s = \pm 2) 
\end{array}\right. .
\end{align}
\end{subequations}
Here, we have truncated the range of displacement $s$ to $|s|\le 2$. This truncation introduces an error of order $\epsilon^3$, which is of the same order as the Trotter error. As we shall see, $\{s\}$ constitute the physical legs of the MPS. The truncation reduces the computational cost by decreasing the physical dimension of the MPS without sacrificing much precision.

By definition (Eq.~\ref{eq:I_def_x}), the phonon influence functional $I$ depends on electron coordinates $\{x\}$. Its dependence on displacements $\{s\}$ is implicit through the relation $s_j = x_{j} - x_{j-1}$. We wish to make the dependence of $I$ on $\{s\}$ explicit. To this end, we employ the translation operator $T(a)$, which translates the phonon state by $a$ lattice spacings. For instance, $T(1)$ translates the phonon state by 1 lattice spacing:
\begin{align}
T(1)(|\phi_1\rangle_1 \otimes |\phi_2 \rangle_2 \cdots  \otimes |\phi_N\rangle_N) 
\nonumber\\
= |\phi_N\rangle_1 \otimes |\phi_1 \rangle_2 \cdots  \otimes |\phi_{N-1}\rangle_N.
\end{align}
Here, $|\phi\rangle_j$ denotes the phonon state on site $j$. $T(a)$ is obtained by successive applications of $T(1)$, i.e. $T(a) = T(1)^a$. It follows that:
\begin{align}
V(x+a) = T(a) V(x) T(-a).
\end{align}
The above relation can be understood as follows. $V(x)$ is the phonon Hamiltonian when the electron resides on site $x$. Translating the phonons by $-a$ effectively puts the electron on $x+a$ from phonons' perspective, thereby producing the Hamiltonian $V(x+a)$.

Using the translation operator, we may trade $V(x_j)$ that appears in Eq.~\eqref{eq:I_def_x} by $V(0)$, namely the phonon Hamiltonian when the electron is located at the lattice origin. We obtain:
\begin{align}
\label{eq:I_def_s}
I(\{s\}) &= \langle \phi| T(x_M)e^{-i\frac{\epsilon}{2}V(0)} T(-x_M) T(x_{M-1})
\nonumber\\
&\quad \times e^{-i\epsilon V(0)} T(-x_{M-1}) \cdots T(x_2)e^{-i\epsilon V(0)}
\nonumber\\
&\quad \times T(-x_1) T(x_0)e^{-i\frac{\epsilon}{2}V(0)}T(x_0)|\phi\rangle
\nonumber\\
& = \langle \phi| e^{-i\frac{\epsilon}{2}V(0)} T(-s_M)e^{-i\epsilon V(0)} \cdots  
\nonumber\\
&\quad \times e^{-i\epsilon V(0)}T(-s_1)e^{-i\frac{\epsilon}{2}V(0)}|\phi\rangle
\nonumber\\
& \equiv \langle \phi| U(s_M) \cdots U(s_2)U(s_1)|\phi\rangle.
\end{align}
In the second equality, we use the composition property of translation operator, $T(-x_j)T(x_{j-1}) = T(-s_j)$, and the fact that the phonon vacuum is translationally invariant, $T(x_M)|\phi\rangle = T(x_0)|\phi\rangle = |\phi\rangle$. In the third line, we define for brevity the unitary operator:
\begin{align}
\label{eq:U_def}
U(s) \equiv e^{-i\frac{\epsilon}{2}V(0)}T(-s)e^{-i\frac{\epsilon}{2}V(0)}.
\end{align}
The phonon influence functional $I$ now explicitly depends on the electron displacements $\{s\}$ thanks to the translation operators. The trick used in deriving Eq.~\eqref{eq:I_def_s} amounts to a change of reference frame reminiscent of the Lee-Low-Pines transformation~\cite{Lee1953}.

Eq.~\eqref{eq:path_integral_s}, Eq.~\eqref{eq:A0_s}, and Eq.~\eqref{eq:I_def_s} are the key results of this section. 

\subsection{Exponentiation of influence functional \label{sec:exponentiation}}

As we have alluded to in Sec.~\ref{sec:intro}, the main strategy for evaluating the path integral Eq.~\eqref{eq:path_integral_s} is approximating the amplitude as a low rank MPS. Specifically, we wish to approximate the amplitude as (Fig.~\ref{fig:sketch}(b)):
\begin{align}
W(\{s\}) & \equiv W_0(\{s\})I(\{s\}) 
\nonumber\\
& \approx W_M(s_M)\cdots W_2(s_2)W_1(s_1).
\end{align}
Here, $W(\{s\})$ is the amplitude of the polaron world line, which is the free amplitude dressed by phonon influence functional. $W_1(s_1)$, $W_2(s_2)$,$\cdots W_M(s_M)$ are $s$-dependent matrices of size $1\times \chi_{M-1}$, $\cdots$, $\chi_2\times \chi_1$, $\chi_1\times 1$. We assume the maximal bond dimension, i.e. the maximum of $\chi_1$, $\chi_2$, $\cdots \chi_{M-1}$ is $\chi$. We may evaluate the propagator as:
\begin{align}
iG^R(k,t) & \approx \sum_{s_M} e^{iks_M} W_M(s_M) \cdots \sum_{s_1} e^{iks_1}W_1(s_1),
\nonumber\\
& = W_M\cdots W_1.
\label{eq:contraction}
\end{align}
The matrix $W_j \equiv \sum_{s_j} e^{iks_j}W_j(s_j)$. The computational complexity for summing each and every $s_j$ in the first line is $O(5M\chi^2)$, where the factor of 5 comes from summing over displacement $s_j$. The computational complexity for the matrix product in the second line is $O(M\chi^2)$.

The problem thus reduces to compressing $W(\{s\}) = W_0(\{s\})I(\{s\})$. Per Eq.~\eqref{eq:I_def_s}, we may view $I$ as an MPS. Its virtual space is the phonon Hilbert space.  Meanwhile, since the free amplitude $W_0$ factorizes (Eq.~\eqref{eq:A0_s}), we may view it as an MPS with bond dimension 1. Since $W$ is the element-wise product of the two MPS, it can be cast as an MPS, too (Fig.~\ref{fig:sketch}c). However, as the bond dimension of $I$ is the phonon Hilbert space dimension, which is exponentially large, direct compression is infeasible with standard MPS algorithms~\cite{Schollwock2011}. 

In what follows, we show that $I(\{s\})$ can be \emph{exponentiated}, $I(\{s\}) = \exp(\lambda^2 C (\{s\}))$, where $C(\{s\})$ is an MPS. Crucially, the bond dimension of $C$ is now $N+2$ (Fig.~\ref{fig:sketch}d). This fact is central to our compression scheme.

Our key observation is that, regarding the product of $U(s)$ operators that appear in $I(\{s\})$ (Eq.~\eqref{eq:I_def_s}) as an $s$-controlled evolution sequence for phonon quantum states, the phonons remain in the manifold of coherent states because (a) the unitary operator $U(s)$ preserves coherent state structure, and (b) the initial state $|\phi\rangle$ is a coherent state. We parametrize the phonon coherent state as: 
\begin{align}
\label{eq:coh_state_def}
|\psi\rangle = e^{\lambda^2 c}|\lambda \bm{z}\rangle.
\end{align}
Here, $c$ is a complex number. $|\lambda \bm{z}\rangle = \exp(\sum_q \lambda z_q b^\dagger_q)|\phi\rangle$. $\bm{z}$ is an $N$ dimensional column vector, whose entry $z_q$ corresponds to the phonon mode with wave vector $q$.  $\lambda$ is the electron-phonon coupling constant. The scaling factors $\lambda,\lambda^2$ are introduced for later convenience.

The evolution of the phonon quantum state due to the $U(s)$ operators is captured by a recursion relation for the coherent state parameters $c$ and $\bm{z}$. We define: 
\begin{align}
\label{eq:c_z_def}
e^{\lambda^2 c_{j-1}}|\lambda \bm{z}_{j-1}\rangle \equiv U(s_{j-1})\cdots U(s_2)U(s_1)|\phi\rangle,
\end{align}
which is the phonon quantum state up to $s_{j-1}$. We now act the next unitary operator $U(s_j)$ on it and ask for the resulted state:
\begin{align}
\label{eq:recursion_def}
& e^{\lambda^2 c_{j}}|\lambda \bm{z}_{j}\rangle = U(s_{j}) (e^{\lambda^2 c_{j-1}}|\lambda \bm{z}_{j-1}\rangle)
\nonumber\\
&\quad = e^{-i\frac{\epsilon}{2}V(0)} T(-s_{j})e^{-i\frac{\epsilon}{2}V(0)} (e^{\lambda^2 c_{j-1}}|\lambda \bm{z}_{j-1}\rangle).
\end{align}
In the second line, we have substituted in the definition of $U(s_{j})$ (Eq.~\ref{eq:U_def}). 

To progress further, we need to determine the action of the operator $\exp(-i\epsilon V(0)/2)$ and $T(-s)$ on a phonon coherent state. The result of acting $\exp(-i\epsilon V(0)/2)$ on a phonon coherent state can be obtained by tedious but straightforward algebra (Appendix~\ref{app:action}). Here, we simply quote the result:
\begin{subequations}
\label{eq:v_act_coh_state}
\begin{align}
e^{-i\frac{\epsilon}{2}V(0)} (e^{\lambda^2 c} |\lambda \bm{z} \rangle )= e^{\lambda^2 (c + \delta + \bm{\beta}^\dagger \bm{z})} | \lambda (\bm{\Gamma} z + \bm{\alpha}) \rangle,
\end{align}
Here, $\bm{\Gamma}$ is a $N\times N$ diagonal matrix, whose rows and columns are labeled by phonon wave vector. Its diagonal entry reads:
\begin{align}
\bm{\Gamma}_{qq} = e^{-i \omega_{q}\epsilon/2}.
\end{align}
$\bm{\alpha}$ and $\bm{\beta}$ are complex $N$-dimensional vectors whose entries are labeled by phonon wave vector as well:
\begin{align}
\bm{\alpha}_q &= \frac{f^\ast_q}{\sqrt{N}\omega_q}(1-e^{-i \omega_q \epsilon/2});
\\
\bm{\beta}_q &= \frac{f^\ast_q}{\sqrt{N}\omega_q}(1-e^{i \omega_q \epsilon /2}) = -(\bm{\Gamma}^{\dagger}\bm{\alpha})_q.
\end{align}
$\delta$ is a complex number:
\begin{align}
\delta = \frac{1}{N}\sum_q \frac{|f_q|^2}{\omega_q}(i\frac{\epsilon}{2} - \frac{1-e^{-i \omega_q\epsilon/2}}{\omega_q}).
\end{align}
\end{subequations}

We see the action of $\exp(-i\epsilon V(0)/2)$ induces an affine transformation on the coherent state parameters:
\begin{align}
c' = c + \delta + \bm{\beta}^\dagger \bm{z},\quad \bm{z}' = \bm{\Gamma} \bm{z} + \bm{\alpha},
\end{align}
where $c'$ and $\bm{z}'$ parametrize the resulted coherent state. We recast the above in matrix form:
\begin{align}
\label{eq:v_transform}
\begin{pmatrix}
c' \\
\bm{z}' \\
1
\end{pmatrix} = \begin{pmatrix}
1 & \bm{\beta}^\dagger & \delta \\
0 & \bm{\Gamma} & \bm{\alpha} \\
0 & 0 & 1
\end{pmatrix} \begin{pmatrix}
c \\
\bm{z} \\
1
\end{pmatrix}.
\end{align}

We now turn to the translation operator $T(-s)$. To determine its action on phonon coherent state, we use:
\begin{align}
T(a)^{-1} b_q T(a) = e^{-iqa} b_q.
\end{align}
It follows that:
\begin{align}
b_q(T(a)|z_q\rangle) = e^{-iqa}z_q T(a)|z_q\rangle,
\end{align}
which implies
\begin{align}
T(a)|z_q\rangle = |e^{-iqa}z_q\rangle.
\end{align}
From the above result, we obtain:
\begin{align}
T(-s) (e^{\lambda^2 c}|\lambda \bm{z}\rangle) = e^{\lambda^2 c} |\lambda \bm{\Delta}(s) \bm{z} \rangle,
\end{align}
where $\bm{\Delta}(s)$ is a diagonal unitary matrix whose diagonal entries are given by:
\begin{align}
\bm{\Delta}(s)_{qq} = e^{iqs}.
\end{align}
The induced linear transformation on $c$ and $\bm{z}$ is given in matrix form:
\begin{align}
\label{eq:t_transform}
\begin{pmatrix}
c' \\
\bm{z}' \\
1
\end{pmatrix} = \begin{pmatrix}
1 & 0 & 0 \\
0 & \bm{\Delta}(s) & 0 \\
0 & 0 & 1
\end{pmatrix} \begin{pmatrix}
c \\
\bm{z} \\
1
\end{pmatrix}.
\end{align}

Applying Eq.~\eqref{eq:v_transform} and Eq.~\eqref{eq:t_transform} to Eq.~\eqref{eq:recursion_def},  we arrive at the following recursion relation for phonon coherent state parameters $c$ and $\bm{z}$ under the action of the operator $U(s_j)$:
\begin{subequations}
\label{eq:recursion}
\begin{align}
\begin{pmatrix}
c_{j} \\
\bm{z}_{j} \\
1
\end{pmatrix} = B(s_j)
\begin{pmatrix}
S_{j-1} \\
\bm{z}_{j-1} \\
1
\end{pmatrix}.
\end{align}
Here, $B(s)$ is a $s$-dependent, $(N+2)\times (N+2)$ matrix:
\begin{align}
B(s) \equiv
\begin{pmatrix}
1 & \bm{\beta}^\dagger & \delta \\
0 & \bm{\Gamma} & \bm{\alpha} \\
0 & 0 & 1
\end{pmatrix}
 \begin{pmatrix}
1 & 0 & 0 \\
0 & \bm{\Delta}(s) & 0 \\
0 & 0 & 1
\end{pmatrix}
\begin{pmatrix}
1 & \bm{\beta}^\dagger & \delta \\
0 & \bm{\Gamma} & \bm{\alpha} \\
0 & 0 & 1
\end{pmatrix}.
\end{align}
The recursion starts with the initial condition:
\begin{align}
\bm{z}_0 = 0,\quad c_0 = 0,
\end{align}
\end{subequations}
which characterizes the phonon vacuum $|\phi\rangle$.  Note all the entries in $B$ are independent of $\lambda$ thanks to the scaling factor introduced in the definition of $\bm{z}$ and $c$ (Eq.~\eqref{eq:coh_state_def}). 

Having derived the recursion relation, we are ready to evaluate $I(\{s\})$. By definition Eq.~\eqref{eq:c_z_def},
\begin{align}
I(\{s\}) = e^{\lambda^2 c_M }\langle \phi|\lambda \bm{z}_M\rangle = e^{\lambda^2 c_M}.
\end{align}
In other words, $I(\{s\})$ is related to the coherent state parameter $c$ at the last step. We find $c_M$ by using the iteration relation:
\begin{align}
\label{eq:C_def}
c_M &=  \begin{pmatrix}
1 & 0 & 0
\end{pmatrix}\begin{pmatrix}
c_M \\
\bm{z}_M \\
1
\end{pmatrix}
\nonumber\\
& = \begin{pmatrix}
1 & 0 & 0
\end{pmatrix}B(s_M) \cdots B(s_2) B(s_1)\begin{pmatrix}
0 \\
0 \\
1
\end{pmatrix}
\nonumber\\
& \equiv C(\{s\}).
\end{align}
In the second line, we have employed the recursion relation Eq.~\eqref{eq:recursion}. In the last line, we have defined the MPS $C(\{s\})$. Substituting the above into the expression for $I$, we arrive at the desired result:
\begin{align}
\label{eq:exponentiation}
I(\{s\}) = e^{\lambda^2 C(\{s\})}.
\end{align}
By definition, $C(\{s\})$ is an MPS with bond dimension $N+2$ since the matrix $B$ is of size $(N+2)\times (N+2)$. Eq.~\eqref{eq:exponentiation} is the key result of this section. 

It is interesting to compare Eq.~\eqref{eq:exponentiation} with more familiar expressions of phonon influence functional. Following the standard path integral treatment of polaron problem~\cite{Feynman1955}, we may write the phonon influence functional as: 
\begin{align}
\label{eq:I_familiar}
I(\{x\}) = e^{-\lambda^2 \int^t_0 dt_1 \int^{t_1}_0 dt_2 D(x(t_1)-x(t_2),t_1-t_2)},
\end{align}
where the integrand of the double time integral is essentially phonon propagtor:
\begin{align}
D(x,t) = \frac{1}{N}\sum_q |f_q|^2 e^{i(qx-\omega_q t)}.
\end{align}
The phonon influence functional, in this form, is not obviously equivalent to the matrix product form Eq.~\eqref{eq:exponentiation}. Nevertheless, the two forms must be equal for any given electron world line. We have verified numerically in Appendix~\ref{app:double_time} that two expressions produce the same result for an ensemble of randomly drawn world lines. 

The purpose of rewriting the well known phonon influence functional expression Eq.~\eqref{eq:I_familiar} as a seemingly more invovled form Eq.~\eqref{eq:exponentiation} is to facilitate the summation of path integral. We want to approximate the amplitude of electron world line $W(\{s\})$ as an MPS to carry out the efficient summation. Eq.~\eqref{eq:I_familiar} does not bear any obvious connection to an MPS, and, therefore, is not immediately amenable to MPS techniques. By contrast, as we shall see momentarily, the exponentiated, matrix product form Eq.~\eqref{eq:exponentiation} allows us to compress $W(\{s\})$ efficiently using standard MPS algorithms.

We stress that rewriting of Eq.~\eqref{eq:I_familiar} as Eq.~\eqref{eq:exponentiation} is \emph{not} about speeding up the evaluation of the phonon influence functional $I(\{s\})$ or the amplitude $W(\{s\})$ for a given world line $\{s\}$. It is to provide a starting point for the next step, namely casting the functional $W(\{s\})$ in the MPS form to facilitate summation, which we explain now.

\subsection{Compression through flow equation \label{sec:compression}}

After exponentiating the phonon influence functional, the amplitude acquires the form, 
\begin{align}
W(\{s\}) = e^{\lambda^2 C(\{s\})}W_0(\{s\}). 
\end{align}
Differentiating $W$ with respect to $\lambda^2$ yields an ordinary differential equation:
\begin{align}
\frac{dW(\{s\})}{d\lambda^2} = C(\{s\}) W(\{s\}).
\label{eq:flow_equation}
\end{align}
The initial condition is $W(\{s\}) = W_0(\{s\})$ at $\lambda^2=0$. Eq.~\eqref{eq:flow_equation} is akin to a flow equation~\cite{Wegner1994} describing the renormalization of polaron dynamics with increasing electron-phonon coupling. 

The flow equation is analogous to the imaginary time Schr\"{o}dinger equation. To make the analogy explicit, we view $W(\{s\})$ as a wave function:
\begin{subequations}
\begin{align}
W(\{s\}) = \langle \{s\}|W\rangle,
\end{align}
and $C(\{s\})$ as the matrix element of a diagonal operator:
\begin{align}
C(\{s\}) = \langle s|\hat{C}|s\rangle.
\end{align}
The flow equation now reads (Fig.~\ref{fig:sketch}e):
\begin{align}
\frac{d|W\rangle}{d\lambda^2} = \hat{C}|W\rangle,
\end{align}
\end{subequations}
with $\lambda^2$ playing the role of imaginary time. $\hat{C}$ can be thought of as a Hamiltonian. Compared to a physical Hamiltonian, $\hat{C}$ has the features that (a) it is diagonal in the $\{s\}$ basis; (b) it is non-Hermitian.

Interpreting $W(\{s\})$ as a wave function immediately points us to the compression method: We solve the flow equation for $W(\{s\})$ variationally by using the MPS as the trial function. In particular, if we fix the maximal bond dimension of the MPS trial function to a designated integer $\chi$, the variational solution is an MPS approximation to the exact $W(\{s\})$ with maximal bond dimension $\chi$. Increasing $\chi$ improves the accuracy of the MPS approximation and hence the accuracy of the sum.

In practice, we solve the flow equation variationally using the established time-dependent variational principle (TDVP) solver~\cite{Haegeman2011,Haegeman2016}. Solving Eq.~\eqref{eq:flow_equation} is the most computationally expensive step in our method. The TDVP algorithm integrates Eq.~\eqref{eq:flow_equation} with respect to $\lambda^2$ step by step. The computational complexity of each step scales approximately as $(c_1 \chi^3d^3 + c_2 \chi^3 d^2 N + c_3 \chi^2  d^3 N^2) M$, where $\chi$ is the maximal bond dimension, $d$ is the physical dimension of the MPS (or, equivalently, the range of $s_j$), $N$ is the number of sites, $M$ is the length of the MPS (or, equivalently, the number of time slices). $c_{1,2,3}$ are constants. In practice, the first two terms dominate over the last, i.e. the main limiting factor is the bond dimension $\chi$. Integrating the flow equation up to $\lambda = 2\sqrt{2}$ with $\chi = 120$, $N=36$, and $M = 400$ takes $\sim 60$ hours on a single core; this is one of the most demanding task carried out in this work.

The physical time $t$ (Eq.~\eqref{eq:G_def}) translates to the number of time slices through the relation $t = M\epsilon$ as we hold the Trotter time $\epsilon$ fixed. Thus, the physical time $t$ constitutes the length direction of the MPS (Fig.~\ref{fig:sketch}b). Late time $t$ corresponds to longer MPS. The amplitude $W(\{s\})$ with different $M$ are independent from each other. We must solve the flow equation for every physical time $t$ of interest, which can be done in parallel.

Once the MPS approximation to $W$ is obtained, it is then summed according to Eq.~\eqref{eq:contraction}. The computational cost for the summation is almost negligible compared to the solution of flow equation. As the momentum $k$ is merely a parameter in the summation, the computational cost for scanning $k$ over the Brillouin zone does not incur significant computational overhead. 

\section{Results \label{sec:results}}

We apply our method to a host of problems in this section. In Sec.~\ref{sec:spectra_1D}, we investigate the spectral function of one-dimensional polaron model with dispersionless and dispersive phonons. In Sec.~\ref{sec:spectra_2D}, we turn to the spectral function of two-dimensional Holstein model. Finally, we simulate the real time diffusing dynamics of a polaron in one and two dimensions in Sec.~\ref{sec:diffusion}.

\subsection{Spectral function in one dimension \label{sec:spectra_1D}} 

\begin{figure*}
\includegraphics[width = 2\columnwidth]{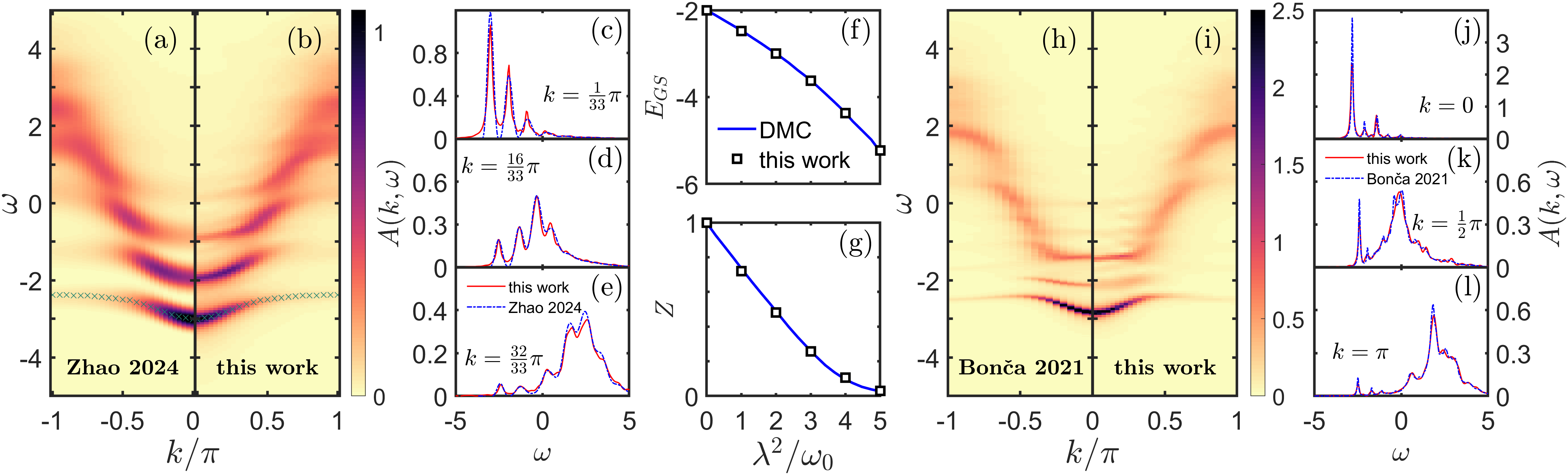}
\caption{Polaron spectral function in one dimension. (a)(b) Spectral function $A(k,\omega)$ for the Holstein model $\omega_q = \omega_0  = 1$, $f_q = 1$, and $\lambda = \sqrt{2}$ calculated with system size $N=16$ and bond dimension $\chi =120$, compared with the result from MPS Chebyshev expansion~\cite{Zhao2023}. Green crosses mark the dispersion of lowest polaron band from a variational method calculation~\cite{Bonca1999}. (c)(d)(e) Constant-$k$ cuts of $A(k,\omega)$ at representative momenta. (f)(g) Polaron ground state energy $E_\mathrm{GS}$ and spectral weight $Z$ for the parameter set $\omega_q = \omega_0 = 1$ and $f_q = 1$ extracted from our results, benchmarked against diagrammatic Monte Carlo data~\cite{Ragni2020}. (h)(i) Spectral function for a model with dispersive phonons. $\omega_q = 1+0.4\cos(q)$, $f_q = 1$, and $\lambda = \sqrt[4]{3.36}$, compared with the variational result~\cite{Bonca2021}. (j)(k)(l) Representative constant-$k$ cuts of $A(k,\omega)$. }
\label{fig:benchmark_1d}
\end{figure*}

We compute the spectral function:
\begin{align}
A(k,\omega) = -\frac{1}{\pi}\mathrm{Im}G^R(k,\omega),
\end{align}
where $G^R(k,\omega)$ is the Fourier transform of $G^R(k,t)$: 
\begin{align}
G^R(k,\omega) = \int^\infty_0 G^R(k,t)e^{i\omega t}dt.
\end{align}

We compute $G^R(k,t)$ up to $t_{max} = 80/W$, where $W = 4$ ($W=8$) is free electron bandwidth in one dimension (two dimensions). The Trotter time $\epsilon = 0.4/W$. The Fourier transform requires the $G^R(k,t)$ data over the entire half time axis. In practice, we have only access to data on a finite interval. We therefore calculate the Fourier transform from time domain data by following the standard procedure~\cite{White2008,Barthel2009}. We extrapolate time domain data from $[0,t_{max}]$ to longer times by linear prediction,  which improves the frequency resolution of the spectral function. As $G^R(k,t)$ typically show persistent oscillations (for instance, Fig.~\ref{fig:time_domain_holstein}a), carrying out the Fourier integral over a finite time interval exhibits artifacts on the frequency axis. For instance, the spectral function may lose the positive definite property. We damp these oscillations by a window function $e^{-\eta t}$ $(\eta\sim O(0.1))$ in the Fourier integral, thereby removing these artifacts.

The window function introduces a small degrees of uncertainty for the Fourier transform in that it broadens the spectral peaks by $\eta$. However, the main features do not change with $\eta$. Introducing $\eta$ to the Fourier transform is mathematically equivalent to introducing a small imaginary part to the Lehmann representation of the propagator.

Fig.~\ref{fig:benchmark_1d}b shows the spectral function $A(k,\omega)$ for the Holstein model~\cite{Holstein1959} on a chain of $N=16$ sites, corresponding to the parameter set $\omega_q = \omega_0 = 1$, $f_q = 1$, and $\lambda = \sqrt{2}$ in Eq.~\eqref{eq:hamiltonian}. Results for other representative parameters are provided in Appendix~\ref{app:holstein}. The maximal bond dimension is $\chi = 120$. The results have converged with respect to both $N$ and $\chi$. 

In our method, $k$ is just a parameter in evaluating the propagator (Eq.~\eqref{eq:contraction}), we therefore can scan $k$ on a much finer grid than what periodic boundary condition allows, namely $\delta k = 2\pi/N$.  It amounts to applying a twisted boundary condition to the electron on finite lattice. The convergence of the spectra with respect to $N$ (Fig.~\ref{fig:time_domain_holstein}) suggests that $N$ has exceeds the size of the phonon cloud that dresses the electron.

We observe idiosyncratic satellite peaks associated with the polaron formation. It can be compared with the result from a previous MPS Chebyshev expansion calculation (Fig.~\ref{fig:benchmark_1d}a). The lowest polaron band matches quantitatively the variational result~\cite{Bonca1999} (Fig.~\ref{fig:benchmark_1d}b, green crosses).

Detailed comparison of $A(k,\omega)$ (Fig.~\ref{fig:benchmark_1d}(c) to (e)) shows quantitative agreement. Small discrepancy in peak width is due to different energy resolutions of these two methods, which are respectively controlled by maximal evolution time $t_{max}$ and the number of Chebyshev moments. 

We extract the ground state energy $E_G$ and the spectral weight $Z$ from our data, which correspond respectively to the position and area of the lowest energy spectral peak at $k=0$ (Fig.~\ref{fig:benchmark_1d}(f)(g)). To this end, we carry out the Fourier transform of $G^R(k=0,t)$ to obtain the spectral function at this momentum point. We choose a small $\eta$ such that the lowest energy peak is narrow and isolated from the rest of the spectrum. We then calculate its area to obtain $Z$, whereas its peak position determines $E_G$. We set $\omega_q = \omega_0 = 1$, $f_q = 1$ and vary $\lambda$. It is in excellent agreement with the imaginary time diagrammatic Monte Carlo result~\cite{Ragni2020}, which attests to the accuracy of our method.   

We then turn to models with dispersive phonons. Our method offers considerable flexibility in this regard because the dispersion $\omega_q$ and form factor $f_q$ enters $C(\{s\})$ as parameters (Eq.~\eqref{eq:exponentiation}). In other words, introducing dispersion or non-local coupling to the phonons do not increase the computational cost for the same simulation time $t$, system size $N$, and bond dimension $\chi$. However, the model with dispersive phonons or non-local coupling may require larger $\chi$ to yield accurate results. Empirically, we find this is not the case for moderate phonon dispersion. Using the same bond dimension as the Holstein model, namely $\chi = 120$, yields sufficiently accurate results. We set $\omega_q = 1+0.4\cos(q)$ and $\lambda = \sqrt[4]{3.36}$, whereas the other parameters are the same as before. Our results (Fig.~\ref{fig:benchmark_1d}i) are in good agreement with a previous variational calculation~\cite{Bonca2021} (Fig.~\ref{fig:benchmark_1d}h). Compared to the Holstein polaron (Fig.~\ref{fig:benchmark_1d}(a)(b)), the lowest polaron band bends downward near $k=\pm \pi$, mirroring the phonon dispersion. More results for this model can be found in Appendix~\ref{app:dispersive}.

The spectral line shape from the two methods (Fig.~\ref{fig:benchmark_1d}(j) to (l)) at representative momenta agree very well, too. The different peak widths again reflect the different energy resolutions. In Appendix~\ref{app:dispersive}, we have also compared the partially integrated spectral weight $\int^\omega_{-\infty} A(k,\omega')d\omega'$, which is less sensitive to this issue, and find quantitative agreement for several parameter sets.

\subsection{Spectral function in two dimensions \label{sec:spectra_2D}}

We turn to the more challenging problem of computing the spectral function in two dimensions, where previous attempts rely on cluster perturbation theory~\cite{Hohenadler2003,Zhao2024}. Going from one dimension to higher dimensions requires minimal modification for our method. As a result, the complexity of our method does not increase significantly when applied to two dimensions. In two dimensional square lattice, the displacement $s$ now takes value from the set $\{|\Delta x|\le 2\} \otimes \{|\Delta y| \le 2\}$. i.e. the electron can hop in both directions. The free electron amplitude $W_0$ is replaced by the one for tight binding model on square lattice. Finally, the formula for the MPS $C(\{s\})$ is generalized by extending the phonon wave vector from one dimensional Brillouin zone to two dimensions.  In practice, we split each variable $s_j$ into two variables $\Delta x_j$ and $\Delta y_j$. This step reduces the physical dimension of MPS at the expense of MPS length.

\begin{figure}
\includegraphics[width = 0.725\columnwidth]{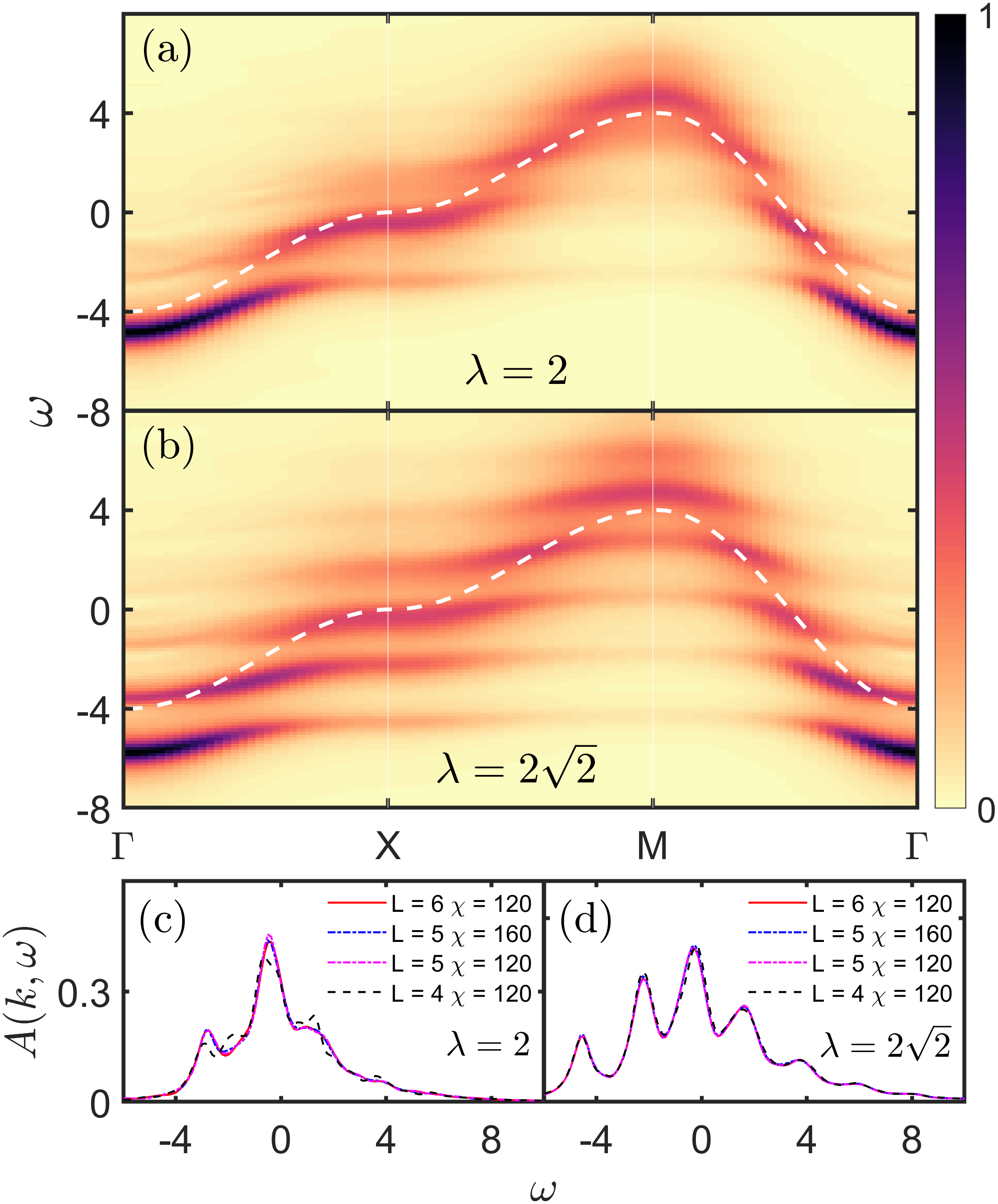}
\caption{Polaron spectral function in two-dimensional square lattice. (a) Spectral function $A(k,\omega)$ along high symmetry directions of the first Brillouin zone for the Holstein model $\omega_q = \omega_0 = 2$, $f_q = 1$, $\lambda = 2$. The system size is $6\times6$. Maximal bond dimension is $\chi = 120$. Dashed line shows the tight binding dispersion. (b) Similar to (a) but for $\lambda = 2\sqrt{2}$. (c)(d) Convergence of the spectral function with respect to the system size and the bond dimension $\chi$ at the X point, which we find empirically exhibits the slowest convergence rate.}
\label{fig:spectra_2d}
\end{figure}

Fig.~\ref{fig:spectra_2d}a shows the spectral function for the Holstein model in two dimensional, $L \times L$ square lattice. We set $\omega_q = \omega_0 = 2$, $f_q = 1$, and $\lambda = 2$. The system size is $L=6$. The maximal bond dimension is $\chi = 120$. We have verified the convergence with respect to both $L$ and $\chi$ (Fig.~\ref{fig:spectra_2d}(c)). Its features resemble that of the one dimensional model. The spectral weight of the lowest band concentrates near $k=0$, whereas the weight of higher bands shifts toward the region with larger $|k|$ in momentum space. Dialing up the coupling strength introduces more bands with weaker dispersion (Fig.~\ref{fig:spectra_2d}(b)(d)). 

\begin{figure}
\includegraphics[width = \columnwidth]{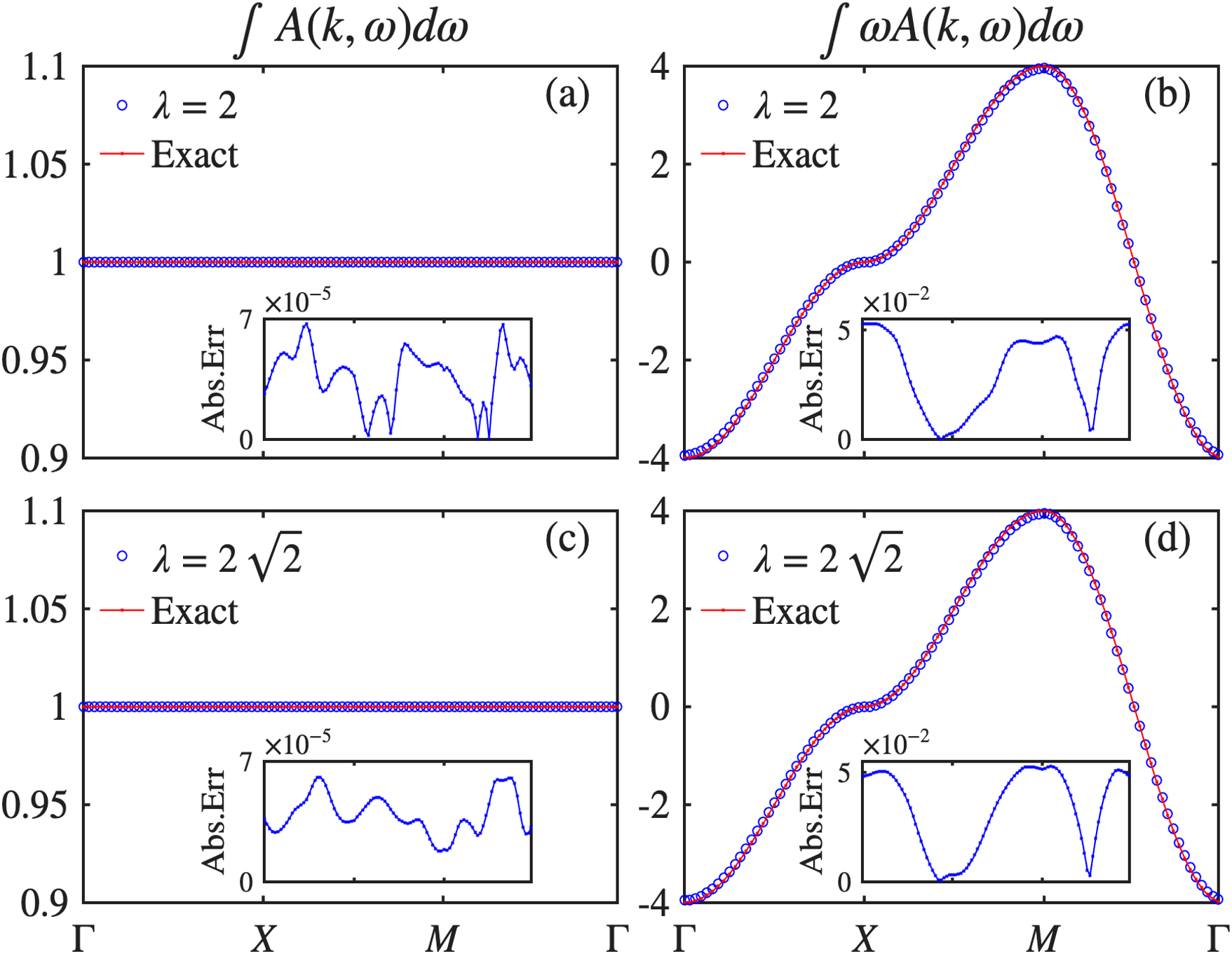}
\caption{Top row: the zeroth (a) and first (b) order moments of the spectral function along the high symmetry directions of the first Brillouin zone for the Holstein model  $\omega_q = \omega_0 = 2$, $f_q = 1$, $\lambda = 2$, corresponding to Fig.~\ref{fig:spectra_2d}a. Blue open circles and red dots are the numerically evaluated value and the exact sum rule, respectively. Insets show the absolute error. Bottom row: similar to the top row but for $\lambda=2\sqrt{2}$, which is to be cross-referenced with Fig.~\ref{fig:spectra_2d}b.}
\label{fig:sum_rule}
\end{figure}

In lieu of previously published, numerically exact result for the spectral function in two dimensions, we benchmark our results by checking the spectral function fulfills the sum rules~\cite{Kornilovitch2002}. The first two sum rules read:
\begin{subequations}
\begin{align}
\int^\infty_{-\infty} A(k,\omega)d\omega &= 1.
\\
\int^\infty_{-\infty} \omega A(k,\omega)d\omega &= E^{(0)}_k.
\end{align}
\end{subequations}
$E^{(0)}_k$ is the free electron dispersion relation. In the present case, $E^{(0)}_k = -2\cos k_x - 2\cos k_y$. Fig.~\ref{fig:sum_rule} shows the calculated zeroth and first moments of the spectral function shown in Fig.~\ref{fig:spectra_2d}(a)\&(b). We find the sum rules are satisfied with the absolute error $\sim O(10^{-5})$ and $\sim O(10^{-2})$ for the zeroth and first moments, respectively. The slight deviation from the exact value is likely due to the Trotter error. The spectral function moments receive contribution over all energy scales, particularly the high energy states. The Trotter discretization has limited impact on low energy features of the spectral function but does affect the high energy features, thereby leading to violation of the sum rules.

\subsection{Diffusion dynamics \label{sec:diffusion}}

\begin{figure}
\includegraphics[width = \columnwidth]{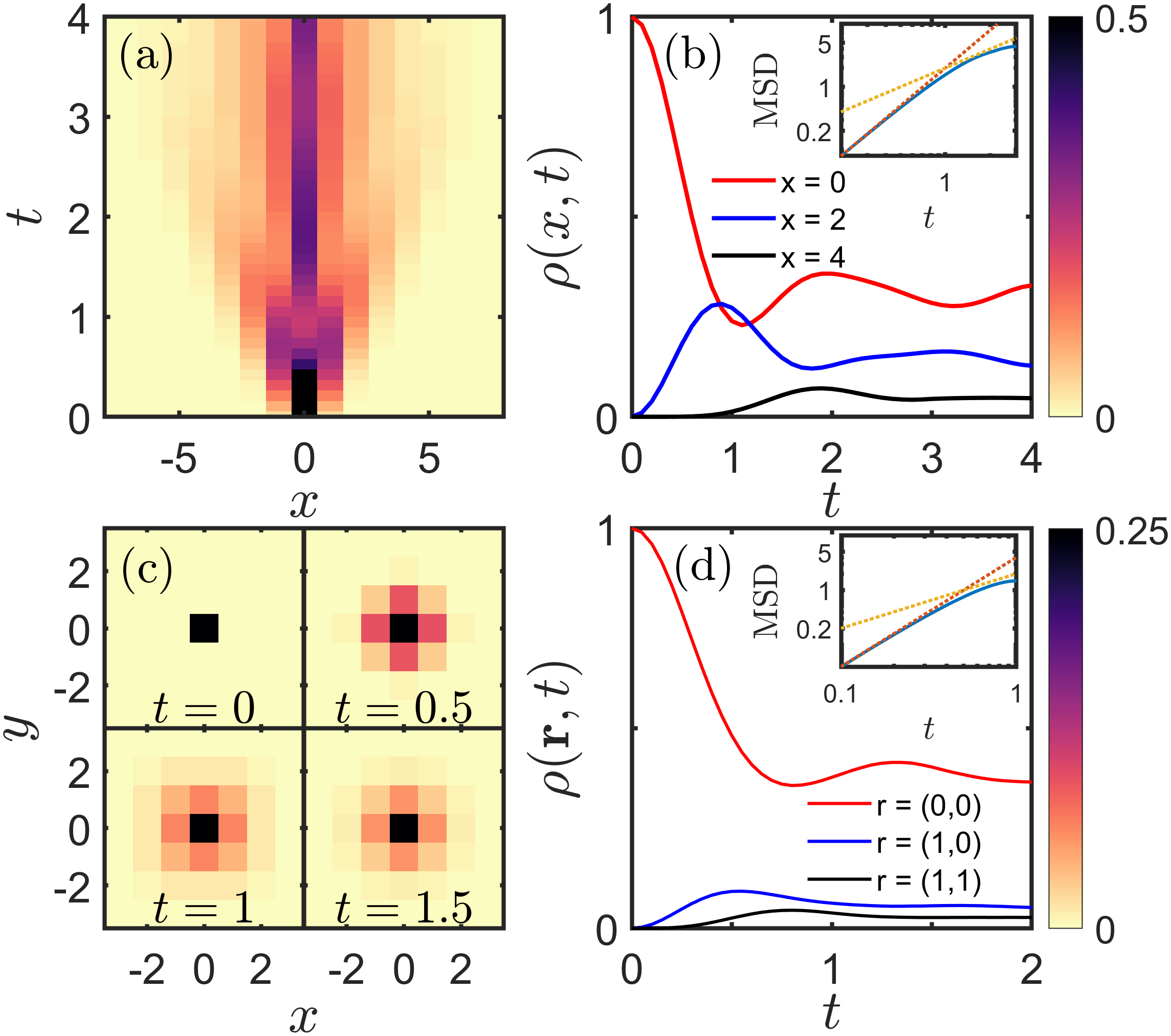}
\caption{Polaron diffusion. (a) Electron density $\rho(x,t)$ as a function of coordinate $x$ and time $t$ for the one-dimensional Holstein model. $\omega_q = \omega_0 = 1$, $f_q = 1$, $\lambda=\sqrt{2}$. The system size $N = 16$. Bond dimension $\chi = 400$. (b) Evolution of $\rho(x,t)$ for selected sites. (c) Snapshots of $\rho(x,t)$ at different times for the two-dimensional square lattice Holstein model. $\omega_q = \omega_0 = 2$, $f_q=1$, $\lambda = 2\sqrt{2}$. We use a $7\times 7$ system with $\chi = 200$. Insets of (b) and (d) show the mean displacement squared (MSD) as a function of time (blue). The ballistic (red) and diffusive (yellow) behaviors are displayed as reference.}
\label{fig:diffusion}
\end{figure}

So far our discussion has been dedicated to the spectral function. We now illustrate the versatility of our method by taking a first pass at another category of problems, namely the diffusion dynamics. Specifically, we plant in an empty lattice an electron at the origin, and evolve it according to the Hamiltonian Eq.~\eqref{eq:hamiltonian}, $|\psi(t)\rangle = e^{-i Ht}c^\dagger_0 |0\rangle$. We calculate the density $\rho(x,t) = \langle \psi(t)|c^\dagger_x c^{\phantom\dagger}_x|\psi(t)\rangle$ by using essentially the same method as the propagator. The corresponding path integral and phonon influence functional in this case are derived in Appendix~\ref{app:diffusion}.

Fig.~\ref{fig:diffusion}a shows the density evolution for the Holstein model in a chain of 16 sites. We set $\omega_q = \omega_0 = 1$, $f_q = 1$, and $\lambda = \sqrt{2}$. It exhibits oscillatory behavior on top of a diffusion-like spreading (Fig.~\ref{fig:diffusion}b). The mean displacement squared (MSD) suggests that it moves ballistically at early time and crosses over to diffusive or sub-diffusive behavior at $t\sim 1$ (Fig.~\ref{fig:diffusion}b, inset). We find similar behavior for the two-dimensional Holstein model on $7\times 7$ square lattice (Fig.~\ref{fig:diffusion}(c)(d)). The crossover occurs earlier due to the larger bandwidth. Note the MSD is only shown for $t<1$ in two dimensions because the electron has reached the boundary beyond this point. Longer simulation time and larger system size are needed to clarify the late time behavior of $\rho(x,t)$ for both cases.

\section{Discussion \label{sec:discussion}}

To summarize, we have shown that the real time Feynman path integral for lattice polaron can be efficiently calculated by compressing the world line amplitude $W(\{s\})$ as an MPS. The success of this approach relies on the fact that $W(\{s\})$ indeed admits an MPS representation with low bond dimension. We understand this fact as follows. The virtual space for the \emph{exact} MPS representation of $W(\{s\})$ is the full phonon Hilbert space (Eq.~\eqref{eq:I_def_s}). Compressing $W(\{s\})$ amounts to searching for a small subspace of dimension $\chi$ that captures the dynamics of the system. We construct this space implicitly but systematically using the TDVP solution of flow equation (Eq.~\eqref{eq:flow_equation}), whereas the previous variational calculations~\cite{Romero1998,Wellein1998,Bonca1999} constructs it explicitly guided by intuition.

Compared with the established numerical method such as Monte Carlo, the main advantage of our method is that it can tolerate complex phase fluctuations in the path integral, thereby allowing for a direct calculation of real time observables. Compared with the variational wave function based approach, our method has the advantage that it can work efficiently in both one and two spatial dimensions, and that it does not require truncating the phonon Fock space. However, our method cannot access the infinite system size limit like the Monte Carlo methods although we have already reached a moderately large system size ($7\times 7$ square lattice). For observables such as the spectral function, we find the result converges rapidly with $N$ for the model parameters considered in this work. To reach larger system size $N$, we could compress the MPS $C(\{s\})$ to a lower bond dimension by using standard singular value decomposition technique, and then use the compressed $C(\{s\})$ for TDVP solver.

Our method is a first step toward a systematic study of polaron dynamics. It can be adapted for Su-Schrieffer-Heeger type models~\cite{Capone1997,Marchand2010} and bipolarons~\cite{Sous2018,Zhang2023}. For the former, a different Trotterization scheme is needed for the path integral; for the latter, the physical dimension of the amplitude is doubled. Furthermore, our preliminary analysis has shown that the method could be extended to the case of thermal phonons~\cite{Bonca2019,Jansen2020} as well as the continuous time limit.

Our method exploits the fact that the phonons stay in the coherent state manifold in the time evolution. For Bose polarons~\cite{Grusdt2024}, it is necessary to extend the manifold to the squeezed coherent states. The influence functional acquires a form different from Eq.~\eqref{eq:exponentiation}. Similar reasoning also applies to other classes of polarons including the Fermi polarons~\cite{Massignan2014} and spin polarons~\cite{Bohrdt2021}. Extending our method to these settings is an interesting open question at the moment. It can be empowered by the tensor cross interpolation method~\cite{Oseledets2010,Oseledets2011,Dolgov2020,NunezFernandez2022,NunezFernandez2025}, thus opening up more possibilities.

\begin{acknowledgments}
We thank Janez Bon\v{c}a and Shuo Yang for sharing data from Refs.~\onlinecite{Bonca2021} and~\onlinecite{Zhao2023}, respectively, and Zi-Xiang Li, Hai-Jun Liao, Tao Shi, Lei Wang, and Tao Xiang for discussions. This work is supported by the National Key Research and Development Program of China (Grants No. 2024YFA1408700, 2021YFA1403800),  by the National Natural Science Foundation of China (Grants No. 12250008, 12188101), and the CAS project for Young Scientists in Basic Research (Grant No. YSBR-059). 

The data that support the findings of this article are openly available~\cite{Data}.
\end{acknowledgments}

\appendix

\section{Proof of Eq.~\eqref{eq:v_act_coh_state} \label{app:action}}

In Sec.~\ref{sec:exponentiation} of the main text, we have derived a recursion relation for the phonon coherent state due to the successive action of the operator $U(s)$. In particular, we have quoted the following result:
\begin{subequations}
\label{app:eq:action}
\begin{align}
e^{-i\frac{\epsilon}{2}V(0)} (e^{\lambda^2 c} |\lambda \bm{z} \rangle )= e^{\lambda^2 (c + \delta + \bm{\beta}^\dagger \bm{z})} | \lambda (\bm{\Gamma} z + \bm{\alpha}) \rangle,
\end{align}
where $\bm{\Gamma}$ is a $N\times N$ diagonal matrix:
\begin{align}
\bm{\Gamma}_{qq} = e^{-i \omega_{q}\epsilon/2}.
\end{align}
$\bm{\alpha}$ and $\bm{\beta}$ are complex $N$-dimensional vectors:
\begin{align}
\bm{\alpha}_q &= \frac{f^\ast_q}{\sqrt{N}\omega_q}(1-e^{-i \omega_q \epsilon/2});
\\
\bm{\beta}_q &= \frac{f^\ast_q}{\sqrt{N}\omega_q}(1-e^{i \omega_q \epsilon /2}) = -(\bm{\Gamma}^{\dagger}\bm{\alpha})_q.
\end{align}
$\delta$ is a complex number:
\begin{align}
\delta = \frac{1}{N}\sum_q \frac{|f_q|^2}{\omega_q}(i\frac{\epsilon}{2} - \frac{1-e^{-i \omega_q\epsilon/2}}{\omega_q}).
\end{align}
\end{subequations}
We prove Eq.~\eqref{app:eq:action} in this appendix.

We note that the action of operator $\exp(-i\epsilon V(0)/2)$ on phonon coherent states factorize for different wave vectors. It is therefore sufficient to consider a single mode with wave vector $q$. The quantity that we need to compute is:
\begin{align}
\exp (-i \frac{\omega_q \epsilon}{2} b^\dagger_q b^{\phantom\dagger}_q + i\frac{\lambda f_q \epsilon}{2\sqrt{N}} b_q + i\frac{\lambda f^\ast_{q} \epsilon}{2\sqrt{N}} b^\dagger_q ) |\lambda z_q\rangle.
\end{align} 
We untangle the operator exponential by the following identity:
\begin{align}
\label{app:eq:untangle}
e^{-Ab^\dagger b + B b + C b^\dagger} = e^{\frac{BC}{A}-\frac{BC}{A^2}(1-e^{-A})} e^{-Ab^\dagger b} 
\nonumber\\
\times e^{C\frac{e^{A}-1}{A}b^\dagger} e^{B\frac{1-e^{-A}}{A}b}.
\end{align}
The proof of this identity is left at the end of this appendix. Setting,
\begin{align}
A = \frac{i\omega_q\epsilon}{2};
\quad
B = i\frac{\lambda f_q \epsilon}{2\sqrt{N}};
\quad
C = i\frac{\lambda f^\ast_q\epsilon}{2\sqrt{N}},
\end{align}
we obtain:
\begin{align}
&\quad\exp (-i \frac{\omega_q \epsilon}{2} b^\dagger_q b^{\phantom\dagger}_q + i\frac{\lambda f_q \epsilon}{2\sqrt{N}} b_q + i\frac{\lambda f^\ast_{q} \epsilon}{2\sqrt{N}} b^\dagger_q ) 
\nonumber\\
&= \exp[ i\frac{\lambda^2|f_q|^2}{2N\omega_q}\epsilon - \frac{\lambda^2|f_q|^2}{N\omega^2_q}(1-e^{-i\frac{\omega_q\epsilon}{2}})]
\nonumber\\
& \times \exp(-i\frac{\omega_q\epsilon}{2} b^\dagger_q b^{\phantom\dagger}_q)\exp[\frac{\lambda f^\ast_q}{\sqrt{N}\omega_q}(e^{i\frac{\omega_q\epsilon}{2}}-1)b^\dagger_q] 
\nonumber\\
&\times \exp[\frac{\lambda f_q}{\sqrt{N}\omega_q}(1-e^{-i\frac{\omega_q\epsilon}{2}})b_q].
\end{align}
Computing its action on coherent state is now straightforward:
\begin{align}
&\quad \exp (-i \frac{\omega_q \epsilon}{2} b^\dagger_q b^{\phantom\dagger}_q + i\frac{\lambda f_q \epsilon}{2\sqrt{N}} b_q + i\frac{\lambda f^\ast_{q} \epsilon}{2\sqrt{N}} b^\dagger_q )  |\lambda z_q\rangle
\nonumber\\
& = \exp[ i\frac{\lambda^2|f_q|^2}{2N\omega_q}\epsilon - \frac{\lambda^2|f_q|^2}{N\omega^2_q}(1-e^{-i\frac{\omega_q\epsilon}{2}})]
\nonumber\\
& \times \exp [\frac{\lambda^2 f_q}{\sqrt{N}\omega_q}(1-e^{-i\frac{\omega_q\epsilon}{2}})z_q]
\nonumber\\
& \times |\lambda e^{-i\frac{\omega_q\epsilon}{2}}z_q+\frac{\lambda f^\ast_q}{\sqrt{N}\omega_q}(1-e^{-i\frac{\omega_q\epsilon}{2}})\rangle.
\end{align}
Collecting the results for different $q$ yield Eq.~\eqref{app:eq:action}.

Eq.~\eqref{app:eq:untangle} is essentially a variation of Feynman's disentangle formula~\cite{Feynman1951}. We prove it as follows. We write:
\begin{align}
e^{Ab^\dagger b}e^{-Ab^\dagger b + B b + C b^\dagger} = \mathcal{T}_u e^{\int^1_0 (Bb (u) + Cb^\dagger(u)) du}.
\end{align}
Here, 
\begin{subequations}
\begin{align}
b(u) &\equiv e^{uAb^\dagger b}be^{-uAb^\dagger b} = e^{-u A}b;
\\
b^\dagger(u) &\equiv e^{uAb^\dagger b}b^\dagger e^{-uAb^\dagger b} = e^{u A}b^\dagger,
\end{align}
\end{subequations}
are in the ``interacting" picture with respect to $Ab^\dagger b$. $\mathcal{T}_u$ denotes the operator ordering with respect to the parameter $u$. 

The right hand side can be evaluated using the Magnus expansion:
\begin{align}
& R.H.S. = e^{\int^1_0 (Bb (u) + Cb^\dagger (u)) du}
\nonumber\\
& \times e^{\frac{1}{2}\int^1_0 du_1 \int^{u_1}_0 du_2 [Bb (u_1) + Cb^\dagger (u_1),Bb  (u_2) + Cb^\dagger (u_2)]}.
\end{align}
The Magnus expansion truncates to second order because the commutator in the second line is a $c$-number. Completing the integral, we obtain:
\begin{align}
R.H.S. = e^{\frac{B}{A}(1-e^{-A})b + \frac{C}{A}(e^A-1)b^\dagger}
\nonumber\\
\times e^{\frac{BC}{A} - \frac{BC}{2A^2}(e^A-e^{-A})}.
\end{align}
We untangle the first term using the Zassenhaus formula:
\begin{align}
R.H.S. = e^{\frac{C}{A}(e^A-1)b^\dagger} e^{\frac{B}{A}(1-e^{-A})b}e^{\frac{BC}{A}-\frac{BC}{A^2}(1-e^{-A})},
\end{align}
which is equivalent to Eq.~\eqref{app:eq:untangle}.

\section{Phonon influence functional as a double time integral \label{app:double_time}}

In this appendix, we discuss the connection of the exponentiated, matrix product expression of the phonon influence functional Eq.~\eqref{eq:exponentiation} to the analytic expression of the phonon influence functional commonly encountered in the standard path integral treatment of polaron problem~\cite{Feynman1955}. 

We first derive the said analytic expression for the phonon influence functional, which is defined as:
\begin{align}
\label{app:eq:I_def}
I(\{x\}) = \langle \phi|e^{-i\frac{\epsilon}{2}V(x_M)}\cdots e^{-i\epsilon V(x_{2})} e^{-i\frac{\epsilon}{2}V(x_0)} |\phi\rangle.
\end{align}
It is convenient for the present purpose to split the operator $V(x)$ into two pieces:
\begin{subequations}
\begin{align}
V(x) = V_0 - \lambda V_1(x),
\end{align}
where
\begin{align}
V_0 &= \sum_q \omega_q b^\dagger_q b^{\phantom\dagger}_q,
\end{align}
is the free phonon Hamiltonian. $V_1$ is the perturbation:
\begin{align}
V_1(x) &= \frac{1}{\sqrt{N}}\sum_q (f_q e^{iqx} b_q + f^\ast_q e^{-iqx} b^\dagger_q).
\end{align}
\end{subequations}

We switch to the interaction picture with respect to $V_0$. $V_1(x)$ now acquires a time dependence in this picture:
\begin{align}
V_{1,int}(x,t) = \frac{1}{\sqrt{N}}\sum_q (f_q e^{i(qx-\omega_q t)} b_q + H.C.).
\end{align}
In the interacting picture, we rewrite Eq.~\eqref{app:eq:I_def} as:
\begin{align}
I(\{x\}) &= \langle \phi |e^{itV_0}\mathcal{T}e^{-i\lambda\int^t_0 V_{1,int}(x(t_1),t_1) dt_1}|\phi\rangle
\nonumber\\
& = \langle \phi |\mathcal{T}e^{-i\lambda\int^t_0 V_{1,int}(x(t_1),t_1) dt_1}|\phi\rangle.
\end{align}
Here, $\mathcal{T}$ stands for time ordering. In the second equality, we have used the fact $V_0|\phi\rangle = 0$.  $x(t_1)$ is the piecewise continuous electron path:
\begin{align}
\label{app:eq:x_t}
x(t_1) = \left\{ \begin{array}{cc}
x_0 & t_1\in [0,\frac{\epsilon}{2}] \\ 
x_{1\le j\le M-1} & t_1\in [\frac{(2j-1)\epsilon}{2}, \frac{(2j+1)\epsilon}{2}] \\
x_M & t_1 \in [\frac{(2M-1)\epsilon}{2},M\epsilon] 
\end{array}\right. .
\end{align}
Here, $j$ labels the time slice. 

As the operator $V_{1,int}$ is a linear combination of $b_q$ and $b^\dagger_q$, we may evaluate the time-ordered expectation value by using the cumulant expansion. We arrive at the familiar expression:
\begin{align}
\label{app:eq:I_standard}
I(\{x\}) = e^{-\lambda^2 \int^t_0dt_1 \int^{t_1}_0dt_2 D(x(t_1)-x(t_2),t_1-t_2)},
\end{align}
where the integrand $D(x_1-x_2,t_1-t_2)$ is related to the phonon propagator:
\begin{align}
& D(x_1-x_2,t_1-t_2) = \langle \phi|V_{1,int}(x_1,t_1)V_{1,int}(x_2,t_2)|\phi\rangle
\nonumber\\
& \quad = \frac{1}{N}\sum_q |f_q|^2 e^{iq(x_1-x_2)-i\omega_q(t_1-t_2)}.
\end{align}

To make the dependence on electron coordinates $\{x\}$ more explicit, we evaluate the double time integral in Eq.~\eqref{app:eq:I_standard}. Substituting Eq.~\eqref{app:eq:x_t} into Eq.~\eqref{app:eq:I_standard}, and carrying out the time integration, we find:
\begin{align}
\label{app:eq:I_double_time}
I(\{x\}) = e^{\lambda^2 C'(\{x\})},
\end{align}
where we have defined the multivaraible function:
\begin{align}
\label{eq:Cprime_def}
C'(\{x\}) & \equiv -\frac{1}{N}\sum_{j_1\ge j_2} \sum_q |f_q|^2 \zeta_{j_1j_2}(q) e^{iq(x_{j_1}-x_{j_2})}
\nonumber\\
& \times e^{-i\omega_q \epsilon(j_1-j_2)}.
\end{align}
$j_1$ and $j_2$ are time slice indices. The auxiliary quantity $\zeta_{j_1j_2}$ is given as follows:
\begin{align}
\zeta_{j_1j_2} = \left\{\begin{array}{cc}
(\frac{\sin\frac{\omega_q\epsilon}{2}}{\omega_q/2})^2 & (M>j_1>j_2>0) \\
\frac{e^{i\omega_q\epsilon/2}-1}{i\omega_q}\frac{\sin\frac{\omega_q\epsilon}{2}}{\omega_q/2} & (M=j_1>j_2>0) \\
\frac{e^{i\omega_q\epsilon/2}-1}{i\omega_q}\frac{\sin\frac{\omega_q\epsilon}{2}}{\omega_q/2} & (M>j_1>j_2 =0) \\
(\frac{e^{i\omega_q\epsilon/2}-1}{i\omega_q})^2 & (j_1 = M, j_2 = 0) \\
\frac{\epsilon}{i\omega_q} - \frac{1-e^{-i\omega_q \epsilon}}{(i\omega_q)^2} & (M>j_1=j_2>0) \\
\frac{\epsilon/2}{i\omega_q} - \frac{1-e^{-i\omega_q \epsilon/2}}{(i\omega_q)^2} & (j_1=j_2=0,M)
\end{array}\right. .
\end{align}

\begin{figure}
\includegraphics[width = 0.8\columnwidth]{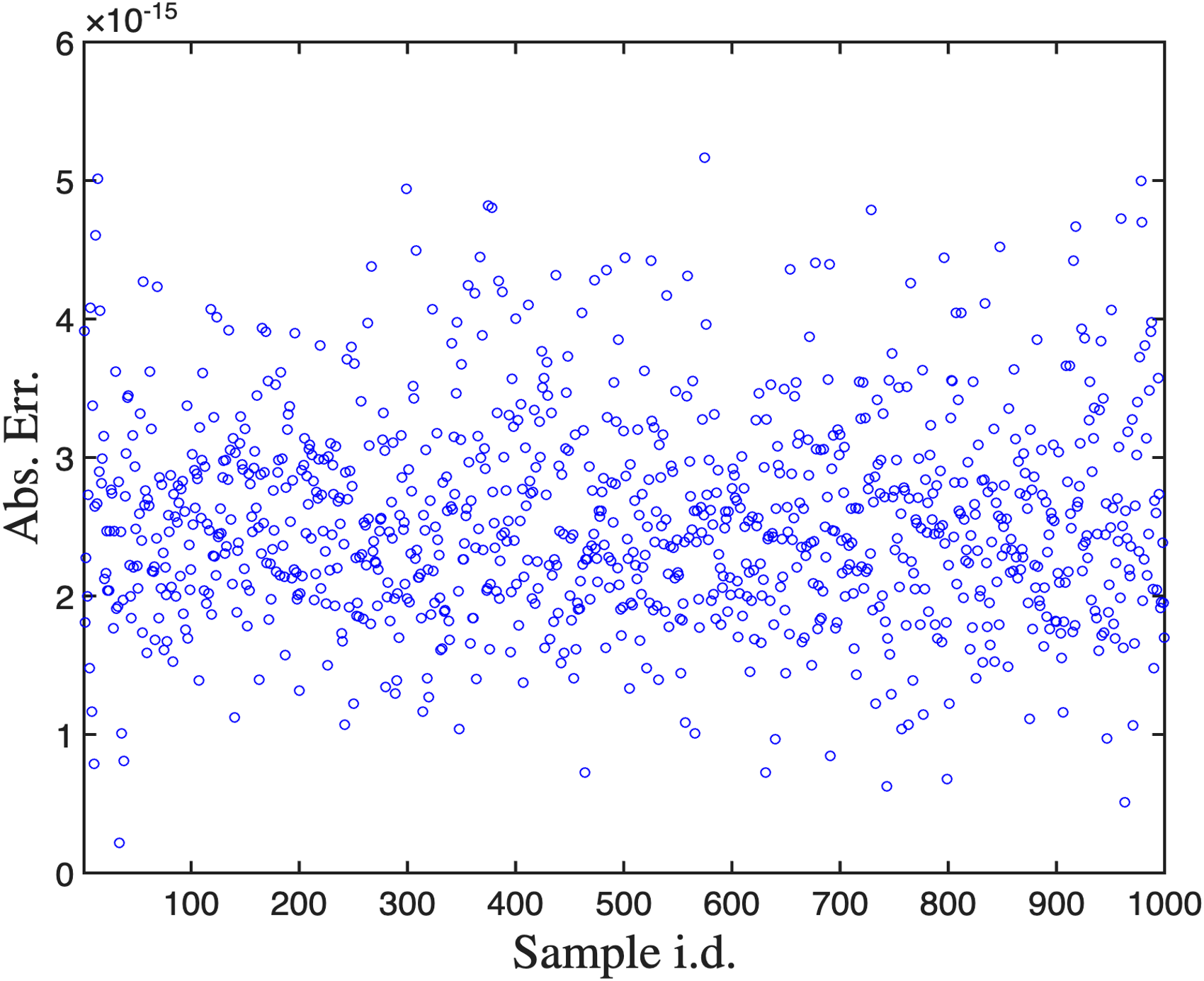}
\caption{Numeric test of the consistency between two different forms of the phonon influence functional over 1000 randomly drawn world lines. The absolute error is defined as $|C(\{s\}) - C'(\{x\})|$, where $C(\{s\})$ (Eq.~\eqref{eq:C_def}) appears in the matrix product form of the influence functional, whereas $C(\{x\})$ appears in the double time integral form. We use a one-dimensional Holstein model with $\omega_q = \omega_0 = 1$, and $f_q = 1$. The system size $N=16$. The number of time slices $M = 20$.}
\label{fig:comparison}
\end{figure}

In the main text (Eq.~\eqref{eq:exponentiation}), we express the phonon influence functional $I(\{s\})$ as:
\begin{align}
I(\{s\}) = e^{\lambda^2 C(\{s\})},
\end{align}
where $C(\{s\})$ is cast in matrix product form (Eq.~\eqref{eq:C_def}). Nevertheless, these two forms must equal. We expect:
\begin{align}
C(\{s\}) = C'(\{x\}),
\end{align}
for any given electron world line. In Fig.~\ref{fig:comparison}, we carry out a numerical test of the above equality over 1000 randomly drawn world lines. We find the difference is on the order of the floating point error. 

\section{Additional data for one-dimensional Holstein model \label{app:holstein}}

\begin{figure*}
\includegraphics[width = 1.5\columnwidth]{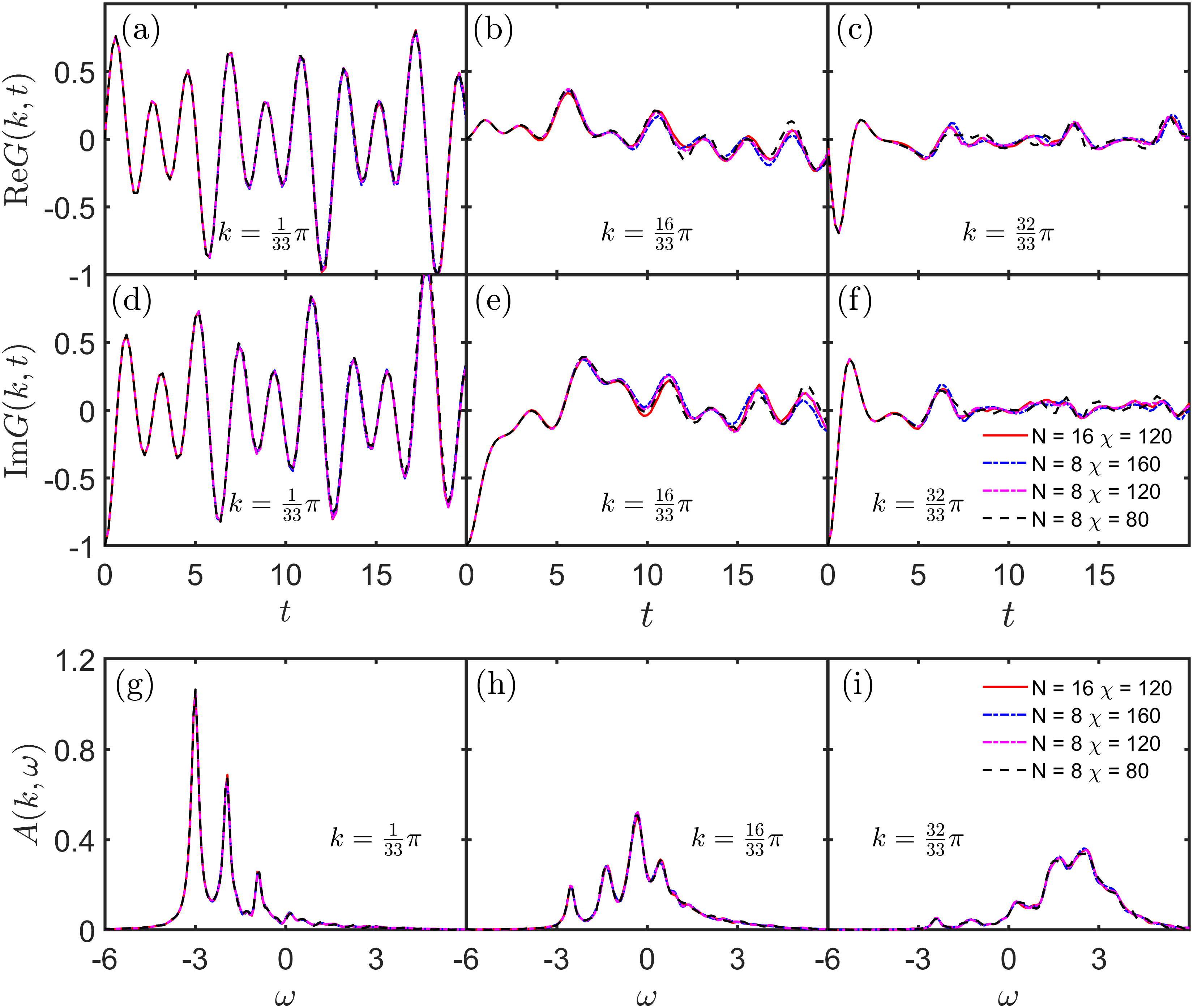}
\caption{Top rows: the real (a$\sim$c) and imaginary (d$\sim$f) part of the electron propagator of the one dimensional Holstein model at three representative wave vectors. The parameters are: $\omega_q = \omega_0 = 1$, $\lambda = \sqrt{2}$, and $f_q = 1$. Bottom row: the corresponding spectral function ($g\sim$i). Data for different system sizes $N$ and bond dimensions $\chi$ are shown.}
\label{fig:time_domain_holstein}
\end{figure*}

Fig.~\ref{fig:time_domain_holstein} shows the data for the electron propagator $G^R(k,t)$ in the time domain at three representative wave vectors. Note these wave vectors are identical to what are shown in Fig.~\ref{fig:benchmark_1d}c$\sim$e in the main text. Both the real (panels a$\sim$c) and the imaginary parts (panels d$\sim$f) are shown. We find that $G^R(k,t)$ at $k\approx 0$ reaches convergence already at low bond dimension $\chi = 80$ and small system size $N=8$. By contrast, $G^R(k,t)$ at $k\approx \pi/2$ and $k\approx \pi$ exhibit slower rate of convergence. 

An interesting feature of the time domain data is that the data at different times exhibit a relatively uniform rate of convergence with respect to the maximal bond dimension $\chi$. We observe a moderate difference, though. $G^R(k,t)$ at early time and some late time intervals reaches convergence relatively early (at $\chi = 80$), whereas, for some specific intervals, there is a discernible difference between data at $\chi = 120$ and $\chi = 160$, indicating that the data at these times have not yet reached full convergence but are fairly close to it. 

The nearly uniform convergence for different times is due to the fact that the physical time corresponds to the length direction of the MPS. It proves advantageous for computing spectral functions (panel g$\sim$i). We find that the spectral functions in fact converge much faster than the time domain data. This behavior is expected: the spectral function at low frequencies are insensitive to small, local changes in the time domain data.

\begin{figure*}
\includegraphics[width = 1.5\columnwidth]{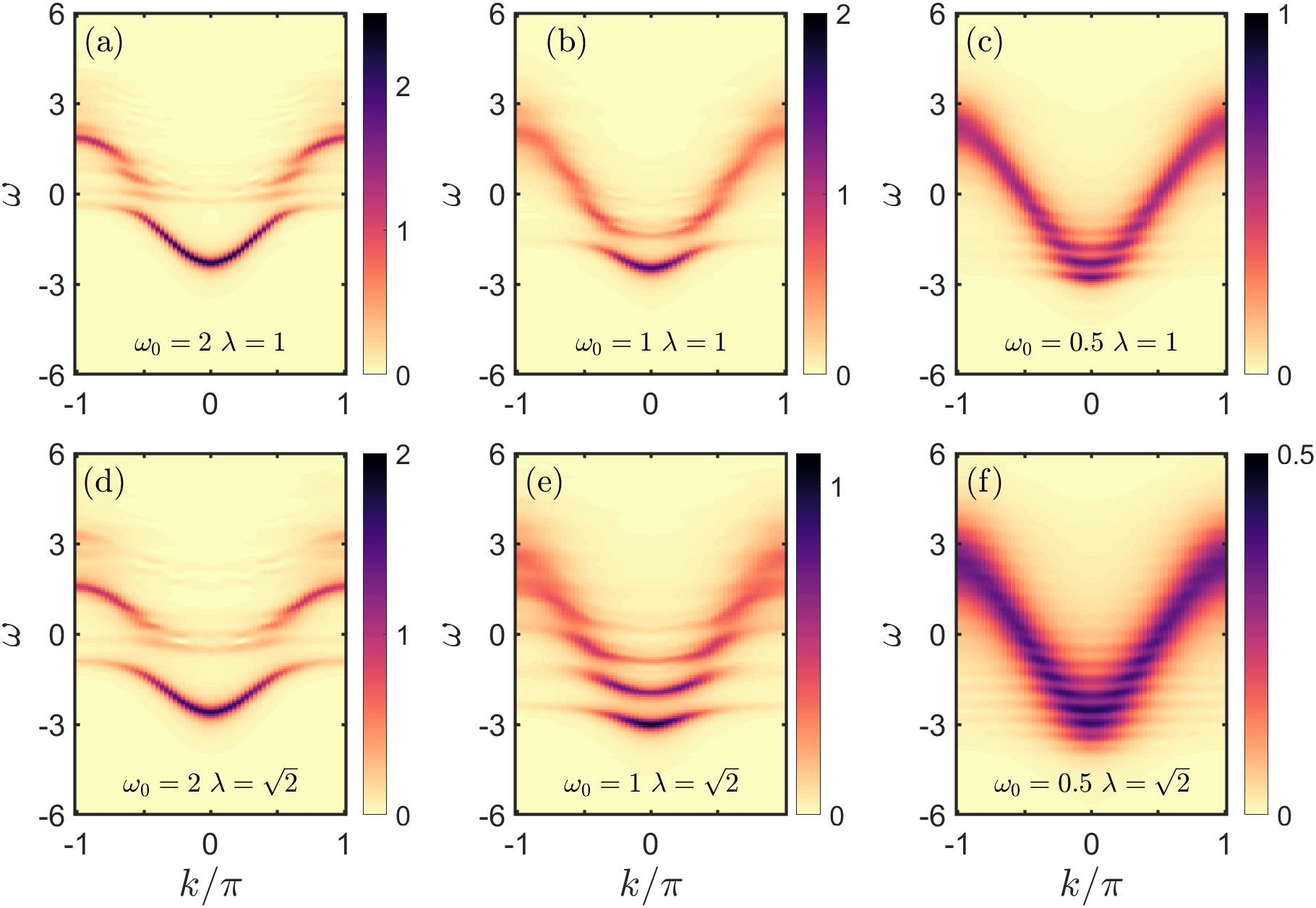}
\caption{Spectral function of one dimension Holstein model. We set $\omega_q = \omega_0$ and $f_q = 1$. The system size $N = 16$. Maximal bond dimension $\chi = 120$.}
\label{fig:holstein_1d}
\end{figure*}

Fig.~\ref{fig:holstein_1d} shows the spectral function of the one-dimensional Holstein model for a few representative parameter sets. We set $\omega_q = \omega_0$ and $f_q = 1$. The system size $N = 16$. The maximal bond dimension $\chi = 120$. The top row (panel a$\sim$c) show the result for $\lambda = 1$ with increasingly softer phonon frequency. The bottom row (panel d$\sim$f) is similar but with stronger coupling constant $\lambda = \sqrt{2}$. The overall trend is that increasing $\lambda$ and/or decreasing $\omega_0$ introduce more satellite peaks, reflecting the fact that the combination $\lambda^2/\omega_0$ controls the effective coupling strength. Interestingly, we find that, for $\omega_0 = 2$, we observe shadow bands with relative small spectral weight between the main bands. We have verified that these are not artifacts. The microscopic origin of these shadow bands are not entirely clear at the moment. 

\section{Additional data for one-dimensional model with dispersive phonons \label{app:dispersive}}

\begin{figure*}
\includegraphics[width = 1.5\columnwidth]{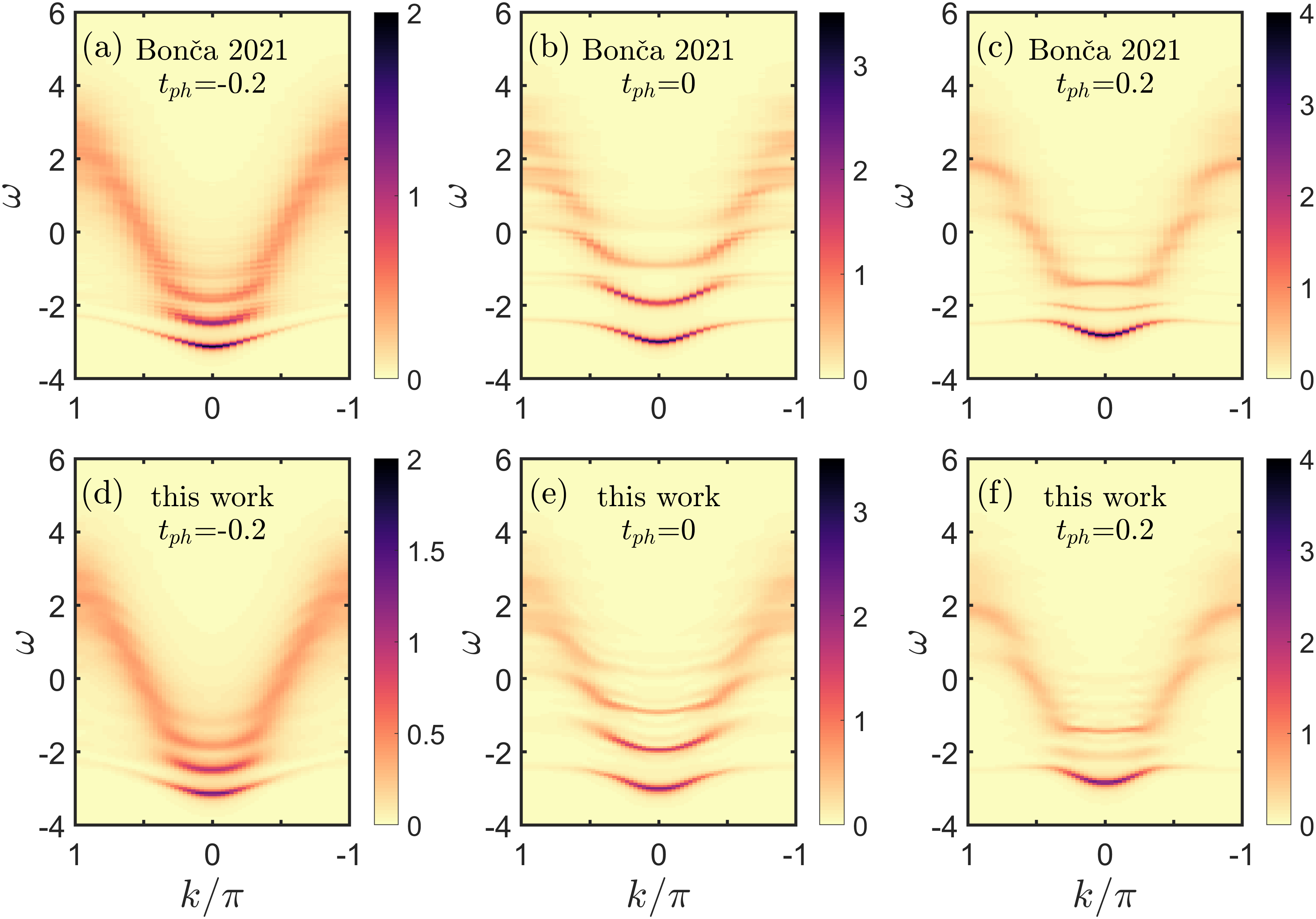}
\caption{Spectral function of one dimension model with dispersive phonons. We set $\omega_q = 1 + 2 t_{ph}\cos q$, $f_q = 1$, and $\lambda = (4-16t^2_{ph})^{1/4}$. The system size $N = 16$. The maximal bond dimension $\chi = 120$. The top row is reproduced from a previous variational calculation~\cite{Bonca2021}, where the bottom row is from this work.}
\label{fig:spectra_dispersive}
\end{figure*}

Fig.~\ref{fig:spectra_dispersive} shows the spectral function for a one-dimensional model with dispersive phonons. Following Ref.~\onlinecite{Bonca2021}, we set $\omega_q = 1+2t_{ph}\cos q$, $f_q = 1$, and $\lambda = (4-16t^2_{ph})^{1/4}$. We use a system of $N = 16$ sites. The maximal bond dimension $\chi = 120$. We find that our results agree well with that of Ref.~\onlinecite{Bonca2021}, which uses a variational construction of low-energy eigenstates. Compared with dispersionless phonons ($t_{ph} = 0$), the satellite peaks for the case of dispersive phonons ($t_{ph}=\pm 0.2$) show narrower spacing in energy. This feature reflects the fact that the phonon frequency minimum $\omega_{min} = 1-2|t_{ph}| = 0.6$ is smaller than the dispersionless case. For $t_{ph}=-0.2$ ($t_{ph} = 0.2$), the lowest polaron band bends upward (downward) near the Brillouin zone boundary, mirroring the phonon dispersion.

\begin{figure*}
\includegraphics[width = 1.5\columnwidth]{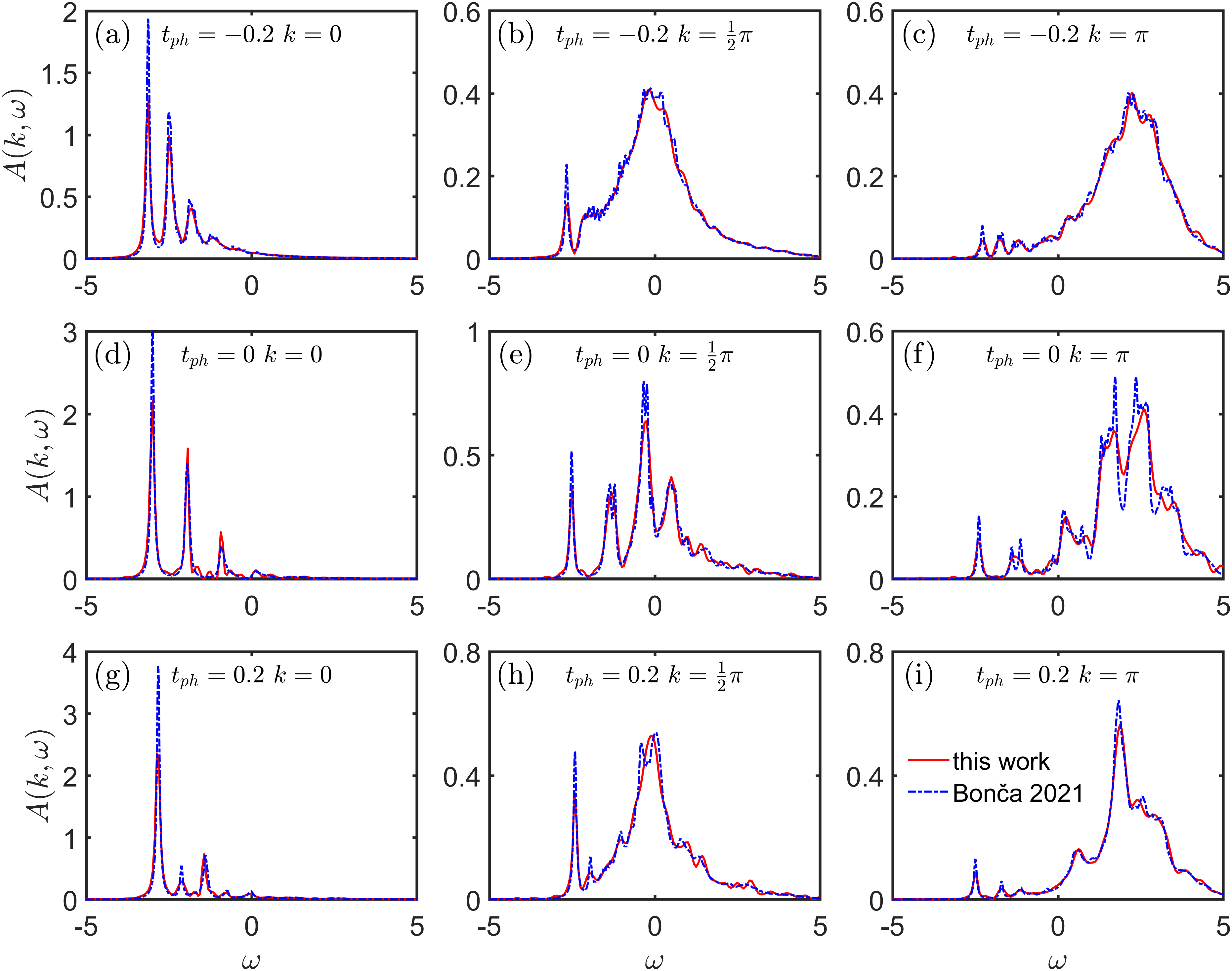}
\caption{Constant $k$-cuts of the spectral function $A(k,\omega)$ for three representative wave vectors $k=0$, $\pi/2$, and $\pi$. The parameter choice is the same as Fig.~\ref{fig:spectra_dispersive}. Both the results from this work (red) and Ref.~\onlinecite{Bonca2021} are shown.}
\label{fig:benchmark_dispersive}
\end{figure*}

We carry out a quantitative comparison of our results with that of Ref.~\onlinecite{Bonca2021} in Fig.~\ref{fig:benchmark_dispersive}. We find good agreement between the two methods. We note that the spectral peaks from Ref.~\onlinecite{Bonca2021} have smaller peak width than ours. This difference is due to the different energy resolution. In the former method, a spectral broadening factor is used, whereas, in the latter, the energy resolution is controlled by the maximal time of evolution.

\begin{figure*}
\includegraphics[width = 1.5\columnwidth]{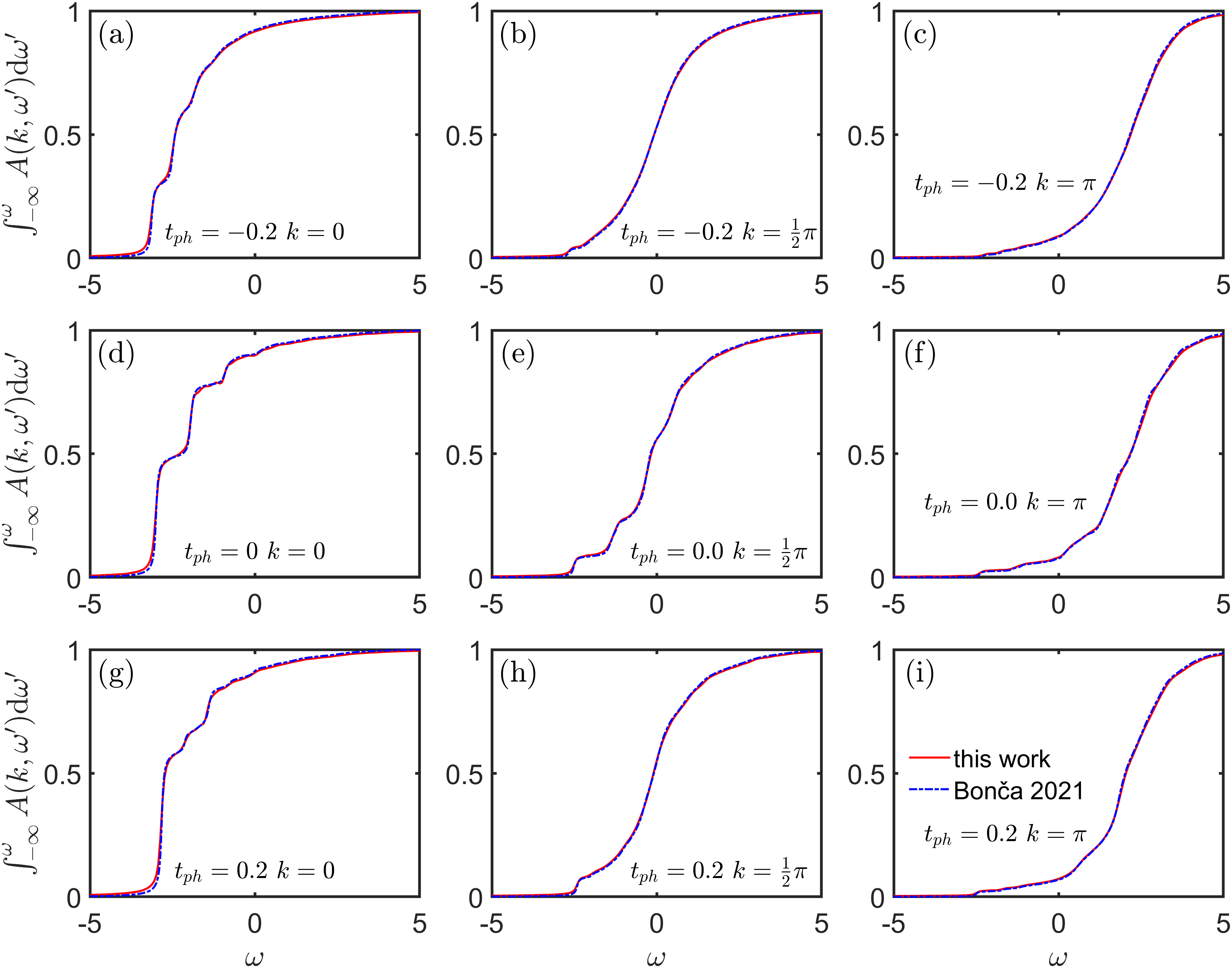}
\caption{Same as Fig.~\ref{fig:benchmark_dispersive} but for the partially integrated spectral function.}
\label{fig:int_spectra}
\end{figure*}

We therefore compare the partially integrated spectral function $\int^\omega_{-\infty} A(k,\omega')d\omega'$. This object is less sensitive to the energy resolution. The steps of the partially integrated spectral function correspond to peaks in $A(k,\omega)$. Fig.~\ref{fig:int_spectra} shows the comparison of the partially integrate spectral function, which shows quantitative agreement between the two methods.

\section{Path integral representation of electron density \label{app:diffusion}}

We consider a nonequilibrium diffusion problem. We place an electron at the origin of an initially empty lattice, $|\psi\rangle = c^\dagger_0|0\rangle$. We monitor the ensuing evolution of the electron density:
\begin{align}
\rho(x,t) = \langle 0|c^{\phantom\dagger}_0 e^{iHt} c^\dagger_x c^{\phantom\dagger}_x e^{-iHt} c^{\dagger}_0|0\rangle,
\end{align}
where $H$ is the Hamiltonian Eq.~\eqref{eq:hamiltonian}. In this section, we derive a path integral representation for $\rho(x,t)$.

First, we trotterize the time evolution and insert the resolution of identity by electron position eigenstates:
\begin{align}
\rho(x,t) = \sum_{\{x^\pm\}} W^\ast_0 (\{x^-\}) W_0(\{x^+\})I(\{x^-\},\{x^+\}).
\end{align}
Here, $\{x^+\}$ and $\{x^-\}$ label the electron position on the forward and backward branch of the Keldysh contour. We impose the boundary condition $x^\pm_0 = 0$ and $x^\pm_M = x$. $W_0(\{x^+\})$ is the amplitude for the electron on the forward branch:
\begin{align}
W_0(\{x^+\}) = \prod^M_{j=1} \langle x^+_{j}|e^{-i\epsilon K}|x^+_{j-1} \rangle.
\end{align}
$W^\ast_0(\{x^-\})$ is the amplitude for the free electron on the backward branch:
\begin{align}
W^\ast_0(\{x^-\}) = \prod^M_{j=1} \langle x^-_{j-1}|e^{i\epsilon K}|x^-_{j} \rangle.
\end{align}
$I(\{x^-\},\{x^+\})$ is the phonon influence functional:
\begin{align}
I(\{x^-\},\{x^+\}) &= \langle \phi | e^{i\frac{\epsilon}{2}V(x^-_0)} e^{i\epsilon V(x^-_1)} \cdots e^{i\epsilon V(x^-_{M-1})} 
\nonumber\\
& \times e^{i\frac{\epsilon}{2}V(x^-_M)} e^{-i\frac{\epsilon}{2}V(x^+_M)} e^{-i\epsilon V(x^+_{M-1})} \cdots 
\nonumber\\
& \times e^{-i\epsilon V(x^+_1)} e^{-i\frac{\epsilon}{2}V(x^+_0)}|\phi \rangle.
\end{align}
Here, $V(x)$ is the same as Eq.~\eqref{eq:V_x}. $|\phi\rangle$ is the phonon vacuum.

Second, we perform a change of variables from the electron position to the displacements, $s^\pm_j = x^\pm_j - x^\pm_{j-1}$. We obtain:
\begin{align}
\rho(x,t) & = \sum_{\{s^\pm\}} \Delta_x (\{s^-\}) \Delta_x(\{s^+\}) W^\ast_0(\{s^-\} )W_0(\{s^+\})
\nonumber\\
&\times  I(\{s^-\},\{s^+\}).
\end{align}
The functional $\Delta_x$ ensures that the electron reaches $x$ at time $t$:
\begin{align}
\Delta_x(\{s^\pm\}) = \delta(x-\sum^M_{j=1}s^\pm_j).
\end{align}
$W_0$ is the amplitude of a free particle defined in Sec.~\ref{sec:path_integral}. The amplitude on the forward and backward branches are complex conjugates. $I$ is the phonon influence functional:
\begin{align}
I(\{s^-\},\{s^+\}) &= \langle \phi | U^\dagger(s^-_1)U^\dagger (s^-_2)\cdots U^\dagger (s^-_M)
\nonumber\\
& \times U(s^+_M) \cdots  U(s^+_2) U(s^+_1) |\phi \rangle
\nonumber\\
& = \langle \phi | U^{-1}(s^-_1)U^{-1} (s^-_2)\cdots U^{-1} (s^-_M)
\nonumber\\
& \times U(s^+_M) \cdots  U(s^+_2) U(s^+_1) |\phi \rangle.
\end{align}
Here, we have used the same trick as the main text, namely using the translation operator to make the dependence on $s$ explicit. $U(s) = e^{-i\frac{\epsilon}{2}V(0)}T(-s)e^{-i\frac{\epsilon}{2}V(0)}$ is the same as before. The second line follows from the unitarity of $U(s)$.

Third, we exponentiate the influence functional. For the time evolution on the forward branch, we use the recursion relation from Sec.~\ref{sec:exponentiation}:
\begin{subequations}
\begin{align}
\begin{pmatrix}
S_{j} \\
z_{j} \\
1
\end{pmatrix} = B(s^+_j) \begin{pmatrix}
S_{j-1} \\
z_{j-1} \\
1
\end{pmatrix}.
\end{align}
The matrix $B(s)$ is defined in Sec.~\ref{sec:exponentiation}. For the time evolution on the backward branch, we exploit the fact that $U^{-1}(s^-)$ is the inverse of $U(s)$:
\begin{align}
\begin{pmatrix}
S_{j} \\
z_{j} \\
1
\end{pmatrix} = B^{-1} (s^{-}_j) \begin{pmatrix}
S_{j-1} \\
z_{j-1} \\
1
\end{pmatrix}.
\end{align}
\end{subequations}
The exponentiated form of the influence functional thus reads:
\begin{align}
I(\{s^-\},\{s^+\}) = e^{\lambda^2 C(\{s^-\},\{s^+\})},
\end{align}
with 
\begin{align}
C(\{s^-\},\{s^+\}) &= \psi_L B^{-1}(s^-_1)B^{-1}(s^-_2)\cdots B^{-1}(s^-_M) 
\nonumber\\
& \times B(s^+_M) \cdots B(s^+_2) B(s^+_1) \psi_R.
\end{align}
The boundary vectors $\psi_L = (1,0,0)$ and $\psi_R = (0,0,1)^T$. 

Since the structure of the phonon influence functional $I$ and the free amplitude $W_0$ are very similar to that of the propagator, we can compress the amplitude $W(\{s^-\},\{s^+\}) \equiv W_0(\{s^-\},\{s^+\}) I(\{s^-\},\{s^+\})$ by using the same method as the main text.  

In the final step, we need to contract $W$ with $\Delta$. It is then necessary to recast $\Delta$ in MPS form. Here, we use the translation matrix $T(x)$:
\begin{align}
T(x)|y\rangle = |x+y\rangle,
\end{align}
where $|x\rangle$ is the electron position eigenstate. It is then easy to see:
\begin{align}
\Delta_x(\{s^+\}) = \langle x|T(s^+_M) \cdots T(s^+_2)T(s^+_1)|0\rangle.
\end{align}
The above is an MPS with bond dimension $N$. The MPS expression for $\Delta_x(\{s^-\})$ can be found in the same vein.

\bibliography{polaron}

\end{document}